\documentclass{jfm}
\usepackage{graphicx}
\usepackage{epstopdf, epsfig}
\usepackage{xcolor}
\usepackage{floatrow}
\usepackage[labelformat = empty,position=top]{subcaption}
\usepackage{caption}
\usepackage[export]{adjustbox}
\usepackage{tabu}
\graphicspath{ {Figure_Latex/} }

\shorttitle{Quasistatic magnetoconvection}
\shortauthor{Z. L. Lim et al.}

\title{Quasistatic magnetoconvection: Heat transport enhancement and boundary layer crossing}

\author{Zi Li Lim\aff{1}, Kai Leong Chong\aff{1}, Guang-Yu Ding\aff{1},
 \and Ke-Qing Xia\aff{1,2}\corresp{\email{xiakq@sustc.edu.cn}}}

\affiliation{\aff{1}Department of Physics, The Chinese University of Hong Kong, Shatin, Hong Kong, China. \aff{2}Department of Mechanics and Aerospace Engineering, Southern University of Science and Technology, Shenzhen, Guangdong 518055, China.}

\begin{document}

\maketitle

\begin{abstract}
We present a numerical study of quasistatic magnetoconvection in a cubic Rayleigh-B\'enard (RB) convection cell subjected to a vertical external magnetic field. For moderate values of the Hartmann number $Ha$ (characterising the strength of the stabilising Lorentz force), we find an enhancement of heat transport (as characterised by the Nusselt number $Nu$). Furthermore, a maximum heat transport enhancement is observed at certain optimal $Ha_{opt}$. The enhanced heat transport may be understood as a result of the increased coherency of the thermal plumes, which are elementary heat carriers of the system.  To our knowledge this is the first time that a heat transfer enhancement by the stabilising Lorentz force in quasistatic magnetoconvection has been observed. We further found that the optimal enhancement may be understood in terms of the crossing between the thermal and the momentum boundary layers (BL) and the fact that temperature fluctuations are maximum near the position where the BLs cross. 
These findings demonstrate that the heat transport enhancement phenomenon in the quasistatic magnetoconvection system belongs to the same universality class of stabilising$-$destabilising ($S$-$D$) turbulent flows as the systems of confined Rayleigh-B\'enard (CRB), rotating Rayleigh-B\'enard (RRB) and double-diffusive convection (DDC). This is further supported by the findings that the heat transport, boundary layer ratio and the temperature fluctuations in magnetoconvection at the boundary layer crossing point are similar to the other three cases.  A second type of boundary layer-crossing is also observed in this work. In the limit of $Re \gg Ha$, the (traditionally defined) viscous boundary $\delta_v$ is found to follow a Prandtl-Blasius-type scaling with the Reynolds number $Re$ and is independent of $Ha$. In the other limit of $Re \ll Ha$, $\delta_v$ exhibits an approximate $\sim Ha^{-1}$ dependence, which has been predicted for a Hartmann boundary layer. Assuming the inertial term in the momentum equation is balanced by both the viscous and Lorentz terms, we derived an expression $\delta_v = H/\sqrt{c_1Re^{0.72} + c_2Ha^2}$ for all values of $Re$ and $Ha$, which fits the obtained viscous boundary layer well.

\end{abstract}


\section{Introduction}

Magnetoconvection -- a fluid convective system subjects to an external magnetic field exists widely in nature. For example, the solar magnetoconvection (\citefullauthor{hurlburt2000solar} \citeyear{hurlburt2000solar}; \citefullauthor{schussler2012solar} \citeyear{schussler2012solar}), evolution of stars (\citefullauthor{macdonald2017apparent} \citeyear{macdonald2017apparent}; \citefullauthor{kitchatinov2001pre} \citeyear{kitchatinov2001pre}), the convection process of liquid batteries (\citefullauthor{shen2016thermal} \citeyear{shen2016thermal}; \citefullauthor{kelley2018fluid} \citeyear{kelley2018fluid}), etc. In many cases, the studies of magnetoconvection are based on the classical Rayleigh-B\'enard (RB) convection (\citefullauthor{ahlers2009rmp} \citeyear{ahlers2009rmp}; \citefullauthor{lohse2010arfm} \citeyear{lohse2010arfm};\citefullauthor{chilla2012epj} \citeyear{ chilla2012epj}; \citefullauthor{xia2013taml} \citeyear{xia2013taml})  -- a fluid layer heated from below and cooled at the top, with a constant temperature difference $\Delta T =T_{bottom}^* - T_{top}^*$, where $T^*$ is the temperature in units. RB convection is characterised by dimensionless control parameters the Rayleigh number $Ra=\alpha g\Delta T/\nu \kappa$ and Prandtl number $Pr=\nu/\kappa$. Here $\alpha$ is the thermal expansion coefficient, $g$ is the gravitational constant, $\nu$ is the kinematic viscosity, and $\kappa$ is the thermal diffusivity. Another control parameter is the aspect ratio $\Gamma = W/H$, where $W$ is the width and $H$ is the height of the system. 

Various studies have been carried out in the past on RB convection in the presence of a magnetic field \citep{burr2002rayleigh}. A particular interest  is RB convection with an applied external horizontal magnetic field \citep{yanagisawa2013convection,tasaka2016regular}. It is observed that under strong magnetic field the convective rolls exhibit quasi-two dimensional flow patterns, which can be understood in terms of linear stability analysis \citep{tasaka2016regular}. For weak magnetic field, the magnetic damping effect is weak, which causes the convection pattern less coherent and hence losing its quasi-two-dimensional character. In a study involving vertical magnetic field, it is reported that the direction of magnetic field affects heat transfer rate. The maximum heat transfer was observed when the vector between gravity and magnetic field differs by $30^{\circ}$. Whereas the heat transfer is minimum when the magnetic field is parallel with gravity (\citefullauthor{naffouti2014three} \citeyear{naffouti2014three}).

Efforts have also been made to extend the Grossmann-Lohse theory \citep{grossmann2000jfm}  for the canonical RB problem into magnetoconvection with an external vertical magnetic field. With inclusions of additional induced equations and parameters such as magnetic Prandtl number and Chandrasekhar number, a new type of scaling behaviour was introduced which depends on magnetic parameters \citep{chakraborty2008scaling}. In another study, in which the induced magnetic field was neglected, four distinct regimes have been identified based on the magnetic field strength and the level of turbulence in the flow (\citefullauthor{zurner2016heat} \citeyear{zurner2016heat}). This type of study corresponds to the so-called quasistatic magnetoconvection, where the magnetic diffusion is sufficiently fast so that fluid flow will not be able to influence (or bend) the magnetic field, which is equivalent to the magnetic field remains constant throughout the whole system (\citefullauthor{cioni2000effect} \citeyear{cioni2000effect}).

A recent development in the studies of convective turbulent flows is the emergence of a new paradigm, which is scalar (heat or salt, for example) transport enhancement by a stabilising force. Three systems have been identified to belong to this class of turbulent flows so far, i.e., confined RB (CRB), rotating RB (RRB), and double diffusive convection (DDC) \citep{chong2017prl}. The physical mechanism leading to the enhanced transport is that the coherency of the coherent structures, i.e. thermal or salt plumes, in these flows is increased when subjected to a stabilising force. It was first discovered, in an experimental and numerical study, that when the aspect ratio $\Gamma$ of a Rayleigh-B\'enard cell becomes sufficiently small, the heat transport efficiency, characterised by the Nusselt number, exhibits an unexpected and rather sharp increase (\citefullauthor{huang2013prl} \citeyear{huang2013prl}). The study further showed that the enhancement can be attributed to the increased coherence of the thermal plumes, i.e. hot/cold plumes are found to be hotter/colder than in the unconfined cases when reaching the opposite plates. A later systematic study by \citefullauthor{chong2015prl} (\citeyear{chong2015prl}) further revealed the existence of an optimal heat transport enhancement, i.e. there exists a particular $\Gamma_{opt}$ for each $Ra$ for which the enhancement is maximum. The  optimal transport enhancements are now understood as a result of the optimal coupling between the suction of hot/fresh fluid and the corresponding scalar fluctuations \citep{chong2017prl}. This optimal coupling comes about when the momentum boundary layer becomes comparable to the thermal boundary layer and because of the temperature/salinity fluctuations reaching maximum at the edge of the thermal boundary. The study by \cite{chong2017prl} also shows that similar heat/salt transport enhancement previously found in the RRB and DDC systems can also be understood in terms of this mechanism. This class of flow is now termed Stabilising-Destabilising flows or $\it S$-$\it D$ flows. This phenomenon in fact has two aspects. The first one is enhancement of scalar transport by a stabilising force, be it drag force due to geometrical confinement in the case of CRB (\citefullauthor{huang2013prl} \citeyear{huang2013prl}; \citefullauthor{chong2015prl} \citeyear{chong2015prl}; \citefullauthor{zwirner2018confined} \citeyear{zwirner2018confined}; \citeauthor{chong2018effect} \citeyear{chong2018effect}), or the Coriolis force in the case of RRB (\citeauthor{zhong2009prandtl} \citeyear{zhong2009prandtl}; \citeauthor{stevens2009prl} \citeyear{stevens2009prl}; \citefullauthor{weiss2016heat} \citeyear{weiss2016heat}), or the stabilising temperature field in the case of DDC (\citefullauthor{yang2016pnas} \citeyear{yang2016pnas}). The second aspect is the existence of an optimal enhancement corresponding to certain value of the control parameter that characterises the stabilizing force, i.e. the aspect ratio for CRB, the Rossby number for RRB and the density ratio for DDC, respectively.

The motivation of the present study is to investigate the effect of Lorentz force, which is a stabilizing force in the RB system, in a systematic way and examine whether magnetoconvection under the quasistatic conditions also exhibits heat transport enhancement and belongs to the same class of $\it S$-$\it D$ flows. Our results show that the quasistatic magnetoconvection under a strong vertical magnetic field indeed belongs to this class of flows, i.e. it exhibits heat transport enhancement within proper region of the parameter space and the observed optimal enhancement can also be understood in terms of momentum boundary layer crossing the thermal boundary layer. We also investigate how the magnetic field influences the large-scale flow in the system. In addition, we  study the behaivor of the velocity boundary layer in the limits of weak and strong magnetic fields, which demonstrates a crossover of the boundary layer from a Prandtl-Blasius type to Hartmann type.

The remaining of this paper is organised as follows. Section 2 provides a brief description about the numerical setup of the study. Section 3 presents results and discussions, with Sec. 3.1 gives an overall visual impression of the effects of magnetic field on turbulent thermal convection by providing snapshots of three-dimensional (3D) temperature, and 2D temperature and velocity fields, respectively. Section 3.2 presents the results on how the global Reynolds and Nusselt numbers responds to the various values of the Hartmann number (proportional to the strength of the applied magnetic field), which shows the existence of heat transport enhancement under moderate magnetic field strength despite the overall flow strength being suppressed by the presence of the Lorentz force. Section 3.3 discusses the behaviour of the momentum (not defined traditionally) and thermal boundary layers and how they are related to the optimal heat transport enhancement. Section 3.4 presents scaling behaviour of the Nusselt number with the Rayleigh number. Section 3.5 discusses how the (traditionally-defined) viscous boundary layer changes from a Prandtl-Blasius type under weak magnetic field to a Hartmann-type under strong magnetic field. We conclude and make some remarks in Sec. 4.

\section{Numerical Setup}

\begin{figure}
    \centering
    \includegraphics[width=3in]{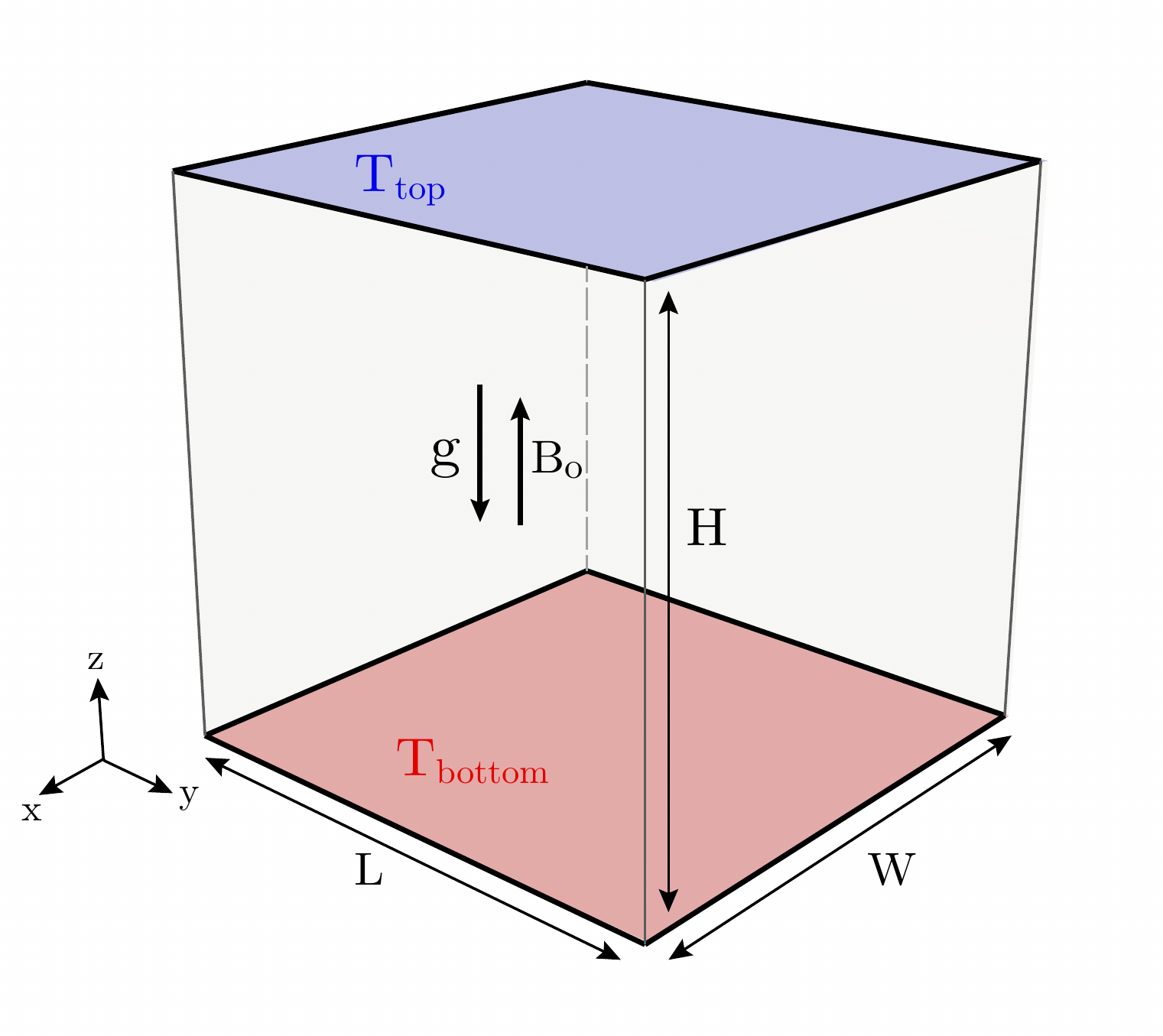}
    \caption{A schematic plot of the Rayleigh-B\'enard cell for numerical simulations. The magnetic field is pointing antiparallel with the gravity.}
    \label{skemplot}
\end{figure}

We perform direct numerical simulation (DNS)  of the three dimensional Navier-Stokes equations subject to a vertical magnetic field and with Boussinessq approximation and the advection-diffusion equation of the temperature in a cubic cell with $L = W = H$ and $\Gamma = W/H = 1$, with $H$ being the height of the cell. The nondimensional equations that describe the velocity field $\mathbf{u}(x,y,z,t)$ and temperature field $T(x,y,z,t)$ is given by: 
\begin{equation}
\partial \mathbf{u}/\partial t + \mathbf{u}\cdot\nabla \mathbf{u} + \nabla p = \sqrt{Pr/Ra}\nabla^{2}\mathbf{u} + T\mathbf{z} + Ha^{2}\sqrt{Pr/Ra}(\mathbf{J} \times \mathbf{B_{0}})
\label{NSeq}
\end{equation}
\begin{equation}
\partial T/\partial t + \mathbf{u}\cdot\nabla T = \sqrt{1/{RaPr}}\nabla^{2}T
\end{equation}
\begin{equation}
\nabla \cdot\mathbf{u} = 0
\label{continuity}
\end{equation}
\begin{equation}
\nabla \cdot\mathbf{J} = 0
\label{eleccon}
\end{equation}
where the bold symbol is in vector form and non-bold symbol is in scalar form. In the above, $\mathbf{J}=\sigma (-\nabla\phi +\mathbf{u}\times \mathbf{B_0})$ is the electric current in the fluid, where $\sigma$ is the electric conductivity of the fluid, and $\phi$ is the electric potential which is a divergence-free component in this system, combined with (\ref{eleccon}) for bounded and insulated system. The external magnetic field is antiparallel to the gravity with a magnitude $\mathbf{B}=B_0\hat{z}$. A dimensionless parameter, the Hartmann number is defined as $Ha = \sqrt{B_0^{2}H^{2}\sigma /\rho_0\nu}= \sqrt{Q}$, where $\rho_0$ is the density of the fluid, and Q is the Chandrasekhar number \citep{chandrasekhar2013hydrodynamic}. As Hartmann number is proportional to the external magnetic field, the magnitude of $Ha$ represents the relative strength of the Lorentz force in the system. Equation (\ref{NSeq}) is the Navier Stokes equation with the Lorentz force included and equation (\ref{continuity}) is the incompressibility condition. The equations are nondimensionlised, with length scale $l_c = H$, free-fall time scale $t_c = \sqrt{H/\beta g\Delta T}$, free-fall velocity $v_c = \sqrt{H\beta g\Delta T}$, and non-dimensional temperature $T = (T^*-T_m)/\Delta T$ with $T_m = (T_{bottom}^*+T_{top}^*)/2$. The dimensionless temperature for the top and the bottom plates are $T_{top} = -0.5$ and $T_{bottom}= 0.5$, respectively. No-slip boundary condition is applied for all six walls. For electric current density, all walls are insulating and it also satisfies the divergence-free condition as shown in equation (\ref{eleccon}). For temperature, the sidewalls are adiabatic, whereas the top and bottom walls are isothermal with value $T_{top}$ and $T_{bottom}$. A sketch of the geometry of the system is shown in figure \ref{skemplot}. 

In contrast with other magnetohydrodynamic DNS studies relevant to astrophysical applications, we consider the case in which the magnetic Reynolds number $R_m = UH/\eta \ll$ 1, where $\eta$ is magnetic diffusivity. In such a case the induced magnetic field due to the current flow diffuse away in a very short time scale, so that it has no influence on the flow. Since the induced magnetic field is neglected, we can assume that the applied external field $\mathbf{B}$ remains constant throughout the convection cell. In other words, the Hartmann number will not change with time. Therefore, we need not to consider the induced equation for magnetic field (\citefullauthor{knaepen2008magnetohydrodynamic} \citeyear{knaepen2008magnetohydrodynamic}).  The DNS was carried out using a multiple resolution version of {\it CUPS}, which has been described in detail in \citefullauthor{CHONG20181045} (\citeyear{CHONG20181045}). The multiple resolution code was used to save the computational cost while achieving the same result in the situation $Pr > 1$. In our simulation, the code is modified to include the Lorentz force term and at the same time ensure that the divergence constraint  in equation (\ref{eleccon}) is satisfied.

The parameter ranges of our study are $Pr = 8$, $Ra = 10^7$ to $10^{10}$, and $0 \leq Ha \leq 800$ for fixed $Ra$ studies; $Ra = 10^5 \sim 10^{10}$, $10 \leq Ha \leq 500$ for fixed $Ha$ studies. As $Pr=8$, the Batchelor length scale is about 3 times smaller than the Kolmogorov length scale. With the multi-resolution scheme,  we use full resolution (velocity grid is equal to temperature grid) for low $Ra$ ($< 1\times10^{7}$) and 1/3 resolution (velocity grid is 1/3 of the temperature grid) for high $Ra$ (\citefullauthor{CHONG20181045} \citeyear{CHONG20181045}). As a check, a 1/3 resolution velocity grid was used in the simulations for low Ra cases as well, and the results are almost the same as those obtained with full resolution. The grid was designed to be denser near boundary layer and coarser in the bulk region in order to resolve boundary layer. Because $Pr$ is larger than 1 in our case, significant saving in computing time is achieved with the multiple resolution method, as solving the velocity field is much more costly than solving the temperature one. The datasets for this study is provided in appendix (table 1 and table 2).

\section{Results and discussion}

To have an overall picture on how magnetic field affects the Rayleigh-B\'enard flow, we first look at the general flow field of the system under different Hartmann. 

\subsection{Visualisation  of the Rayleigh-B\'enard flow in the presence of a vertical magnetic field}

\begin{figure}
    \centering
    \includegraphics[width=5.4in]{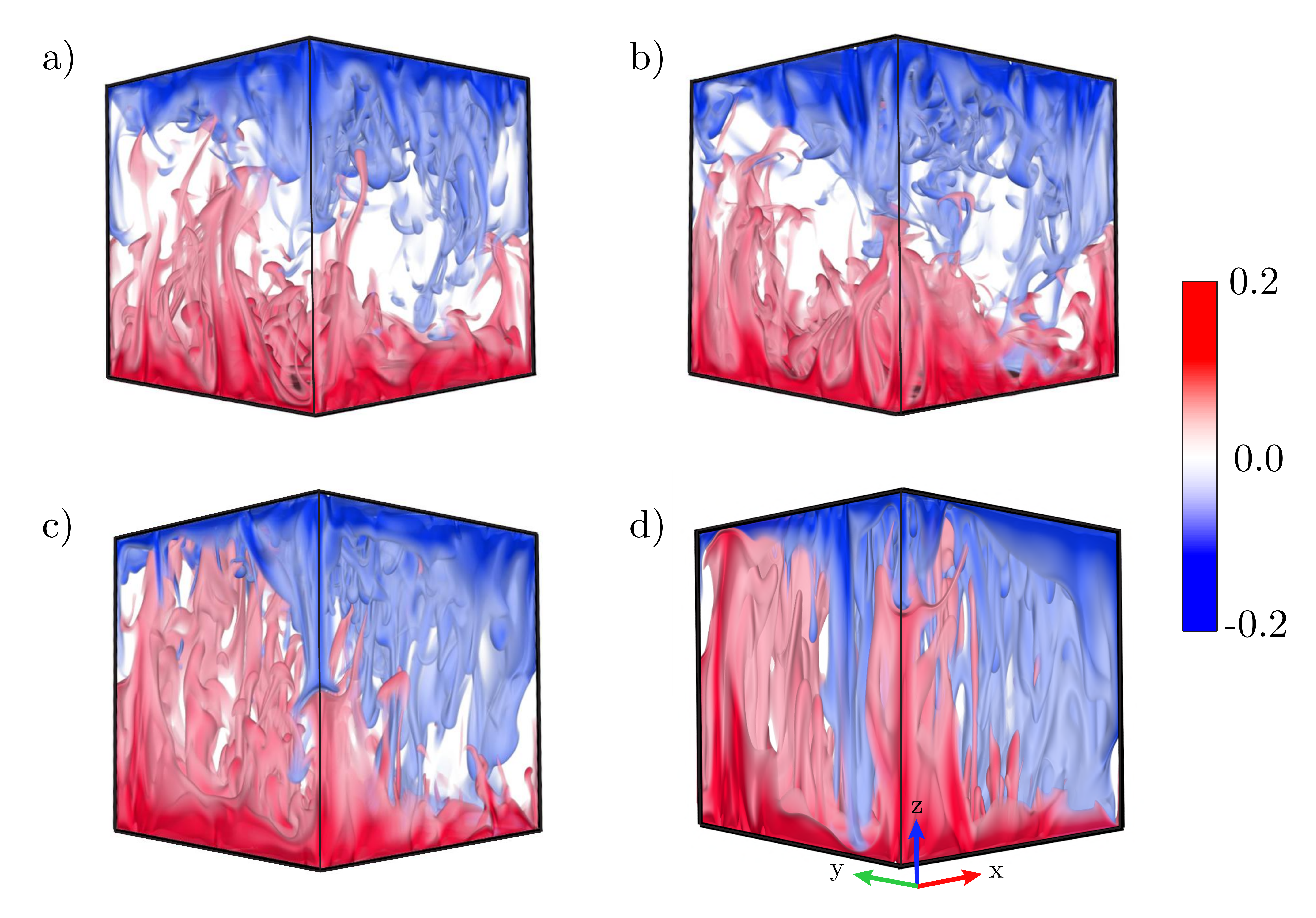}
    \caption{Three dimensional snapshots of the temperature field under different Hartmann numebr at $Ra = 10^{9}$. (a) - (d) refers to the cases $Ha = 10, 50, 200$ and 500 respectively. Temperature scale is shown by the colour bar.  Note how the plume morphology changes as $Ha$ increases.}
    \label{fig:3D}
\end{figure}

\begin{figure}
\centering
\includegraphics[width=5.4in]{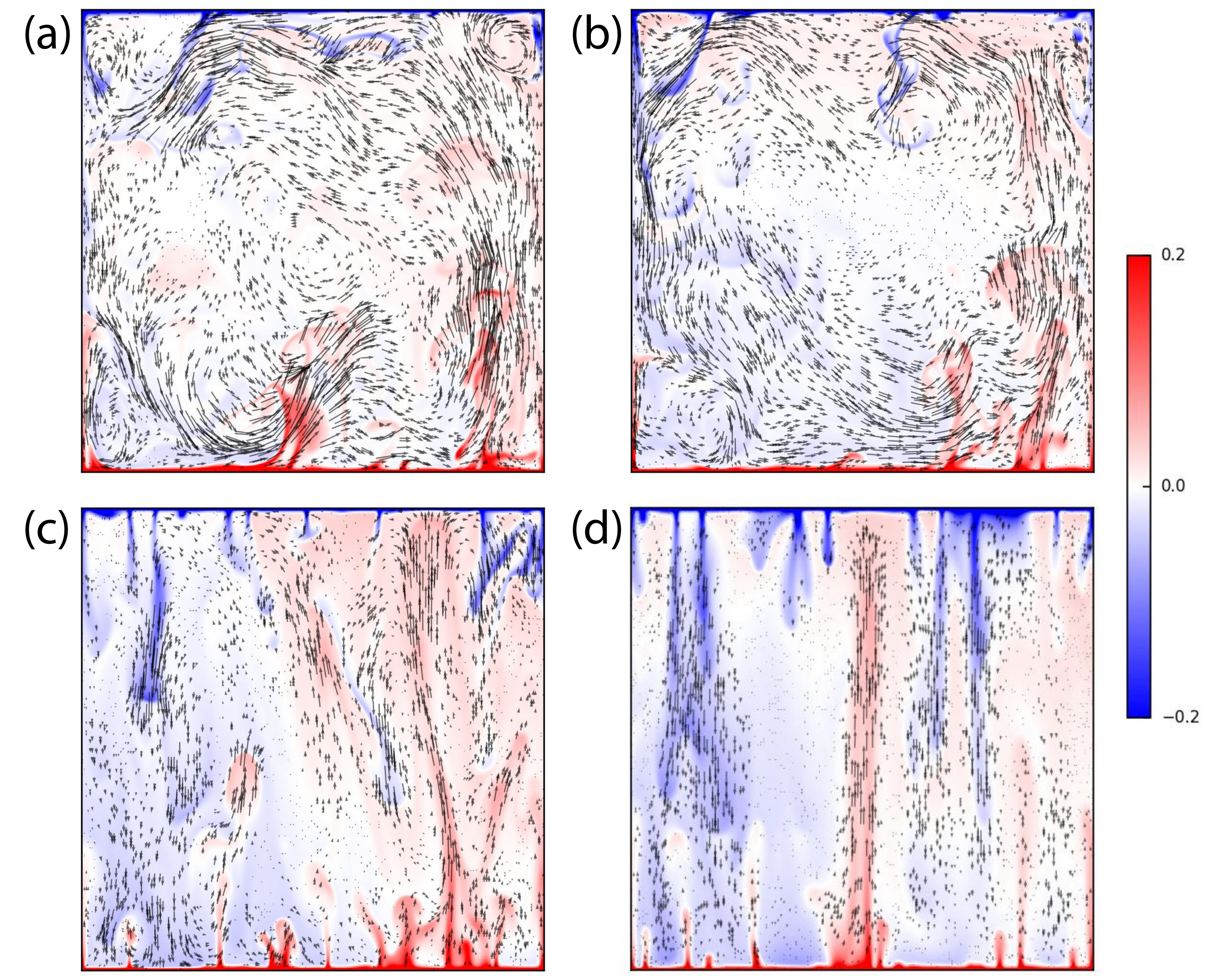}
\caption{Two-dimensional cuts of the instantaneous velocity (arrows) and temperature (colour) fields at vertical mid-plane (y= 0.5) of the flow field at $Ra = 10^{9}$ for different Hartmann numbers. (a) - (d) refers to the cases $Ha = 10, 50, 200$ and 500 respectively. The arrows indicate the flow velocity, with their length represents the velocity magnitude. The colour bar indicates the temperature scale.}
\label{fig:2D}
\end{figure}

Figure \ref{fig:3D} shows  three-dimensional snapshots of the temperature field, with the value given in the scale bar. For the case of $Ha = 10$ and $Ha = 50$, we can clearly see the large-scale circulation (LSC) flowing in the diagonal direction (\citefullauthor{kadanoff2001pt} \citeyear{kadanoff2001pt}; \citefullauthor{xi2004jfm} \citeyear{xi2004jfm}). Based on the figures \ref{fig:3D}(a) and (b) it is difficult to distinguish them from the cases when the magnetic field is absent ($Ha=0$). This is easy to understand since magnetic field suppresses mainly the small-scale flows, the behaviour of the large-scale convective flow  does not change much. In these cases, the plumes are flowing in the direction of the LSC. When the magnetic field becomes sufficiently large, e.g. the case with $Ha = 200$, the plume morphology and behaviour start to change. We can see from figure \ref{fig:3D}(c) that the depth of plume penetration from both the top and bottom plates increases dramatically, which means that the plumes become much larger and more coherent. It is also seen that some of the plumes extend the entire height of the cell. The plumes become more organised, although they do not follow LSC entirely. When the magnetic field increases further, the picture becomes different again as seen in figure \ref{fig:3D}(d). First of all, it is clear that the LSC is completely suppressed. There exist a large number of columnar plumes that extend vertically between the top and the bottom plates. The morphology of cold and hot pillars look very different from the classical mushroom-like plumes. They are narrow and have a clear separation. To understand the columnar plumes, we look at the effect of Lorentz term in (\ref{NSeq}). The vertical magnetic field in the Lorentz term suppresses the horizontal motion of the fluid only. Therefore, in the case of extremely large $Ha$, the fluid would move only in vertical direction, which gives rise to the formation of columnar plumes. This phenomenon was observed in RRB convection (\citefullauthor{stevens2013heat} \citeyear{stevens2013heat}) and in severely-confined RB convection \citep{chong2016jfm} as well. 


To examine the flow behaviour in more detail, we show in figure \ref{fig:2D} vertical cuts of the instantaneous velocity and temperature fields. Although not much difference is observed between the $Ha =10$ and $Ha = 50$ cases, a close examination of the length of the arrows indicates that the magnitude of the flow is decreased with the increasing of the Lorentz force. Also, the large-scale circulation becomes less robust. As Hartmann number increases further to $Ha = 200$, it becomes difficult to ascertain whether the LSC still exist, as the plumes are seen to move mostly in the vertical direction. One may conclude that the LSC has already broken down. In the extreme case of $Ha = 500$, only columnar plumes that extend between the top and bottom plates are observed, and it is clear that the LSC does not exist any more at this very large $Ha$ value, as is already seen from the 3D plot. From the above examples of the instantaneous temperature and velocity fields, we obtain a sense of the role of Lorentz force, which acts to suppress the flow field. To quantify the effect of Lorentz force in RB convection, in the next section we look at the effect of the magnetic field on the time-averaged global quantities, i.e. the Reynolds number and the Nusselt number.

\subsection{Reynolds number and Nusselt number under different $Ha$}

In this section we examine the properties of the two global response parameters, Nusselt number and Reynolds number. For Nusselt number, it is calculated using heat flux across the horizontal plane, which is given by $Nu_h = \left \langle (RaPr)^{1/2}u_zT -\partial T/\partial z  \right \rangle_{x,y,t}$, where $\left \langle \cdot \right \rangle_{x,y,t}$ is the average over a horizontal plane and over time. $Nu$ is then obtained by averaging $Nu_h$ over all horizontal planes. In addition, $Nu$ can also be caculated from the exact relation based on global thermal dissipation, which is given by $Nu_{T} = \left \langle \epsilon_T \right \rangle (RaPr)^{1/2}$. Because of additional Lorentz term in equation (\ref{NSeq}), additional Joule heating effect will contribute to dissipation. The $Nu$ calculated from global viscous dissipation and Joule heating are given by $Nu_{v+b} = 1 + \left \langle \epsilon_v \right \rangle (RaPr)^{1/2} +\left \langle \epsilon_b \right \rangle Ha^2(RaPr)^{1/2}$. $\left \langle \cdot \right \rangle$ denotes time- and volume-average, while the thermal dissipation, viscous dissipation and dissipation from Joule heating are   $\epsilon_T = (RaPr)^{-1/2}\sum\nolimits_{i}(\partial T/\partial x_i)^2$, $\epsilon_v = (Ra/Pr)^{-1/2}\sum\nolimits_{i}\sum\nolimits_{j}(1/2)(\partial u_i/\partial x_j+\partial u_j/\partial x_i)^2$ and $\epsilon_b = (Ra/Pr)^{-1/2}\sum\nolimits_{i}J_i^2$ respectively. The error for Nusselt is estimated by the standard deviation between $Nu$, $Nu_T$ and $Nu_{v+b}$. For Reynolds number, it is defined by $Re =\sqrt{\left \langle \mathbf{u}^2\right \rangle (Ra/Pr)}$. The $Re$ so-defined provides a measure of the overall strength of the flow field.

\begin{figure}
    \centering
    \includegraphics[width=3.5in]{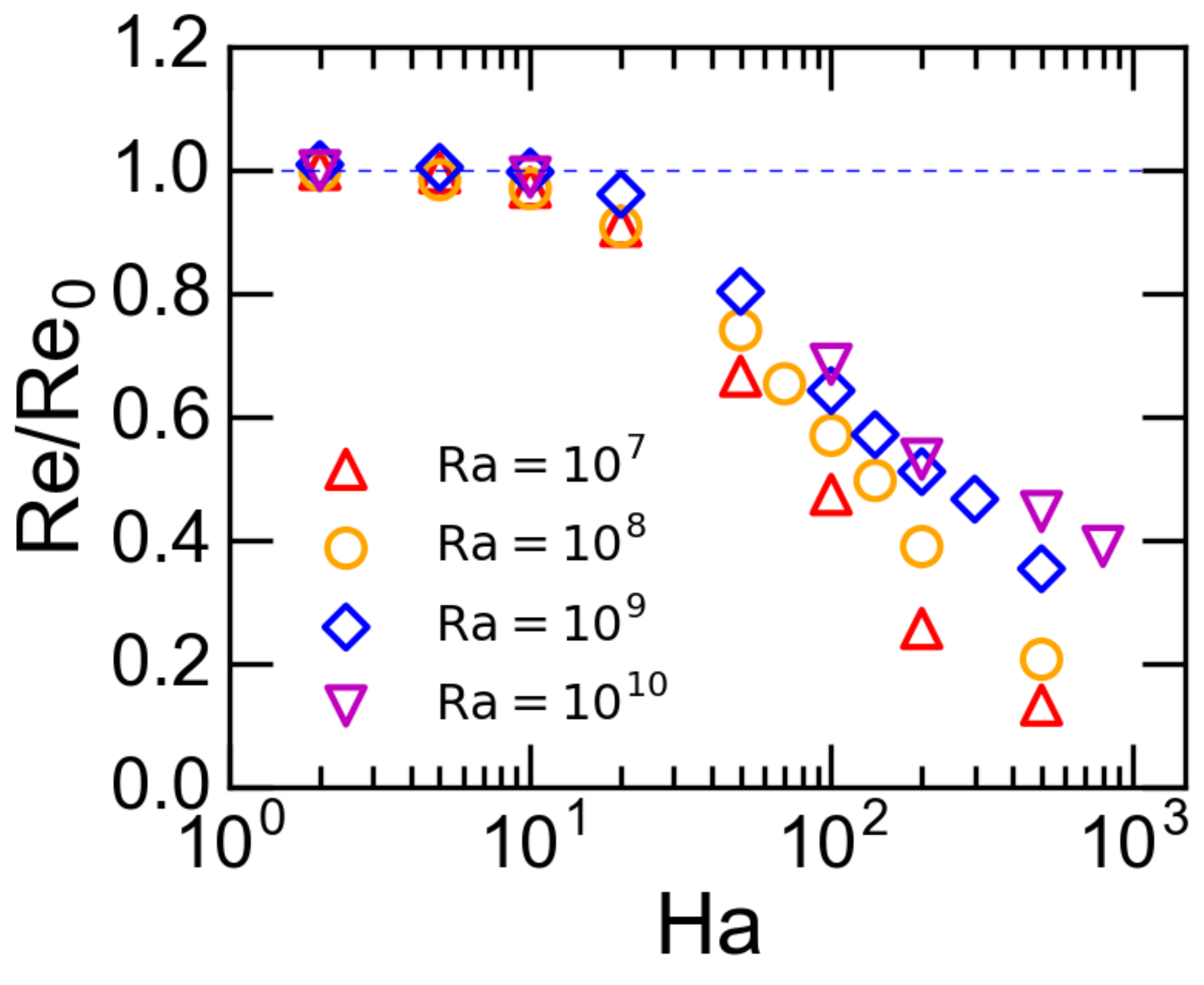}
    \caption{Normalised Reynolds number $Re/Re_0$ versus $Ha$, where $Re_{0}$ is the value obtained at $Ha = 0$}
    \label{Re}
\end{figure}

Figure \ref{Re} plots $Re$ (normalised by its zero-field value) versus $Ha$ for four cases of $Ra$. In all the cases, the value of $Re$ remains almost constant for $Ha \leq 10$ and after that Re decreases monotonically. This is expected since under small Ha, the effect of Lorentz force is too small to alter the flow dynamics of the system. As $Ha$ increases, the Lorentz effect becomes sufficiently strong to suppress the fluid flow. We also see that for the same value of $Ha$, the suppression is more severe for smaller values of $Ra$.


\begin{figure}
\centering
\includegraphics[width=5.4in]{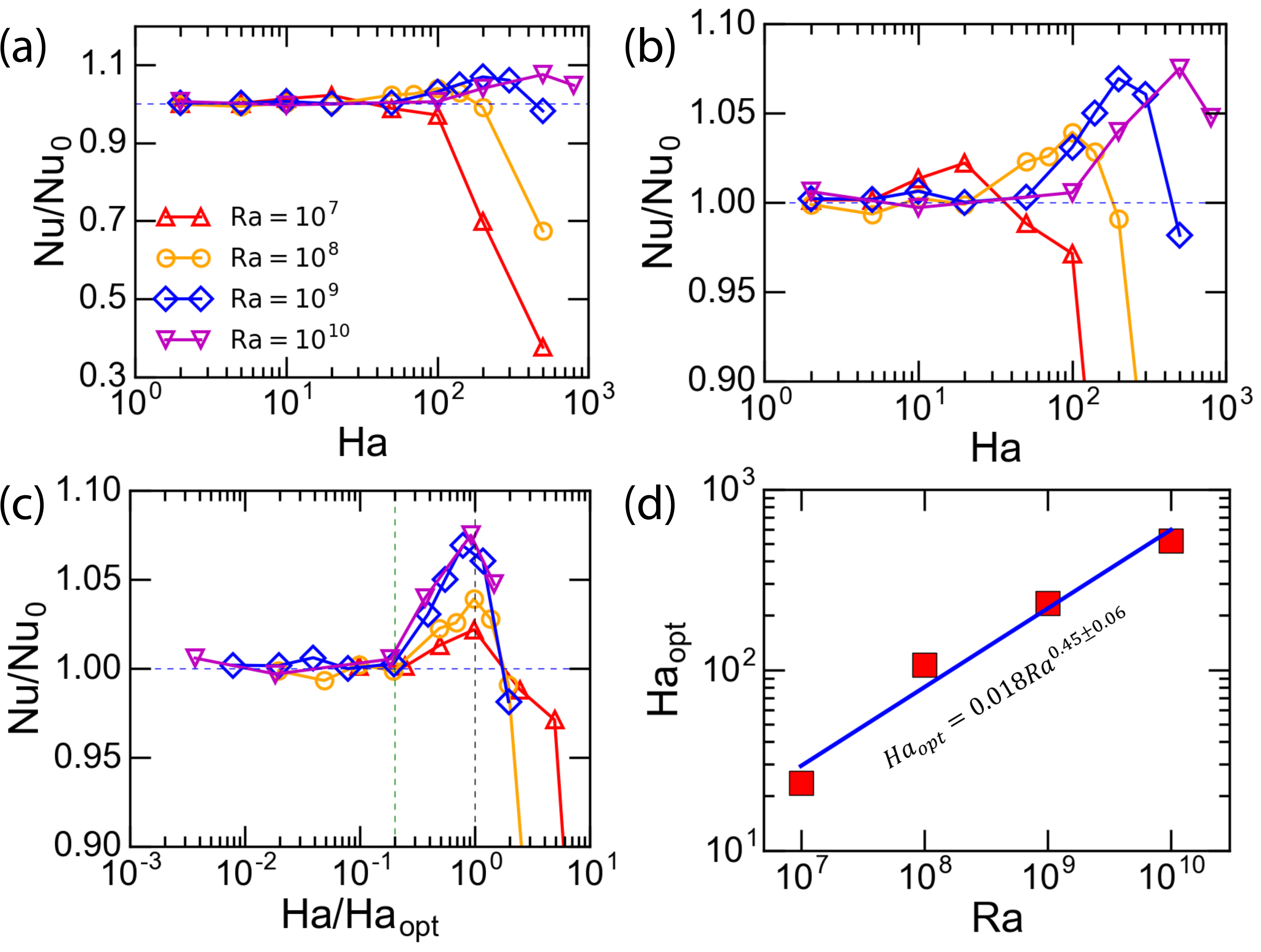}
\caption{Nusselt number behavior and the optimal heat transport enhancement. (a) Normalized  $Nu$ vs. $Ha$, where $Nu_{0}$ is the value obtained at $Ha = 0$. (b) A magnified view of (a). (c) Normalised $Nu$ vs. $Ha/Ha_{opt}$, where $Ha_{opt}$ is where the maximum heat transport occurs. The green dash line indicates $Ha = 0.2Ha_{opt}$ for the onset of heat transfer enhancement and the black dash line indicates $Ha = Ha_{opt}$. (d) $Ha_{opt}$ vs. $Ra$, where a power law fit produces $Ha_{opt} = 0.018Ha^{0.45 \pm 0.06}$.}
\label{Nubehavior}
\end{figure}

Although the flow strength is suppressed as $Ha$ increases, the heat transport behaves very differently. We plot the normalised $Nu$ against $Ha$ in figure \ref{Nubehavior}(a) and an enlarged portion in figure \ref{Nubehavior}(b),  for the same four $Ra$ cases as in the $Re$ plot. It is seen that the detailed behaviour for each $Ra$ may be slightly different, but for all values of $Ra$, the overall trend is the same, i.e., $Nu$ remains unchanged in the beginning and starts to increase as $Ha$ is further increased, it reaches a maximum value at certain $Ha$ (denoted as $Ha_{opt}$). After attaining maximum enhancement, $Nu$ starts to drop sharply. The peak heights are different as well, suggesting that the heat transport enhancement is more prominent for larger $Ra$.
The onset $Ha$ value for enhancement is also seen to be different for different $Ra$. However, if we normalise $Ha$ by the $Ra$-dependent $Ha_{opt}$, then all the curves for different $Ra$ collapse quite well, which is shown in figure \ref{Nubehavior}(c). It is also seen from the figure that the onset Hartmann number for heat transfer enhancement is the same ($Ha = 0.2Ha_{opt}$) for all values of $Ra$, suggesting that $Ha_{opt}$ is a characteristic quantity for transport enhancement. Moreover, we find that the $Ra$-dependence of $Ha_{opt}$ may be described by a power law, $Ha_{opt} = 0.018Ra^{0.45 \pm 0.06}$, as shown in figure \ref{Nubehavior}(d).  
 
Although both are global quantities, the Nusselt and the Reynolds number behave very differently. For example, $Re$ changes monotonically and it starts to decrease at a single transition point of $Ha \approx 10$ for all $Ra$ cases. For $Nu$, on the other hand, the $Ha$ values for the onset of enhancement and for the optimal enhancement are all $Ra$-dependent. This suggests a decoupling between momentum and heat transport in the system, similar to those seen in the other three systems, i.e.  CRB, RRB and DDC. 
For example, in the $Ra = 10^{10}$ and $Ha= 500$ case, $Re$ is already suppressed by $\sim 50\%$ compared to its $Ha = 0$ value, while the maximum enhancement for Nu occurs for this case. 
To understand the heat transport behaviour, such as the existence of the optimal Nu enhancement, we next examine the properties of the thermal and momentum boundary layers and the interplay between the stabilizing and destabilizing forces in the quasistatic magnetoconvection system. 

\subsection{The role of the thermal and stress boundary layers in optimal heat transport enhancement}
\begin{figure}
\centering
\includegraphics[width=5.4in]{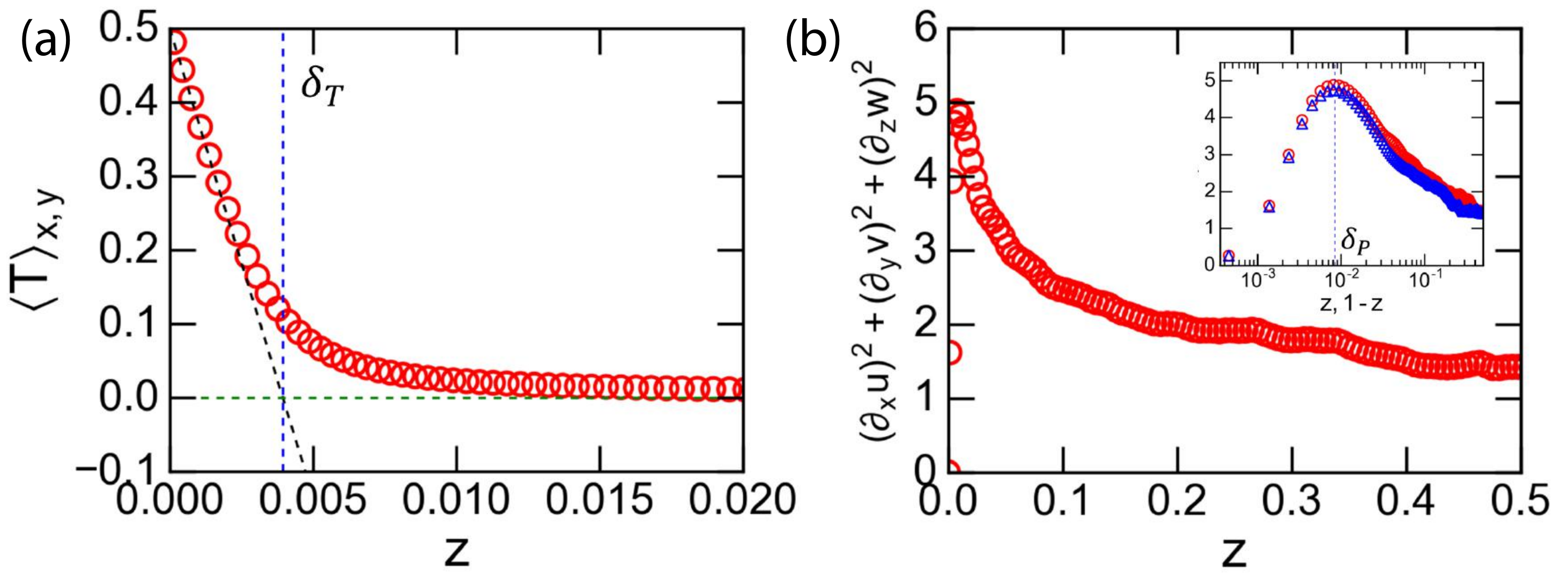}
\caption{(a) Mean temperature profile. The dashed lines define the thickness of the thermal boundary $\delta _T$. (b) Stress profile; the main figure is for height $0\leq z \leq 0.5$, and the inset shows data for both $0\leq z \leq 0.5$ (red circles) and $0.5\leq z \leq 1$ (blue triangles). The dashed vertical line indicates the location of stress boundary layer $\delta_P$.  Both (a) and (b) are for the case of $Ra = 10^{10}$ and $Ha = 10$.}
\label{fig:profile}
\end{figure}

\begin{figure}[tb]\centering
\includegraphics[width=5.4in]{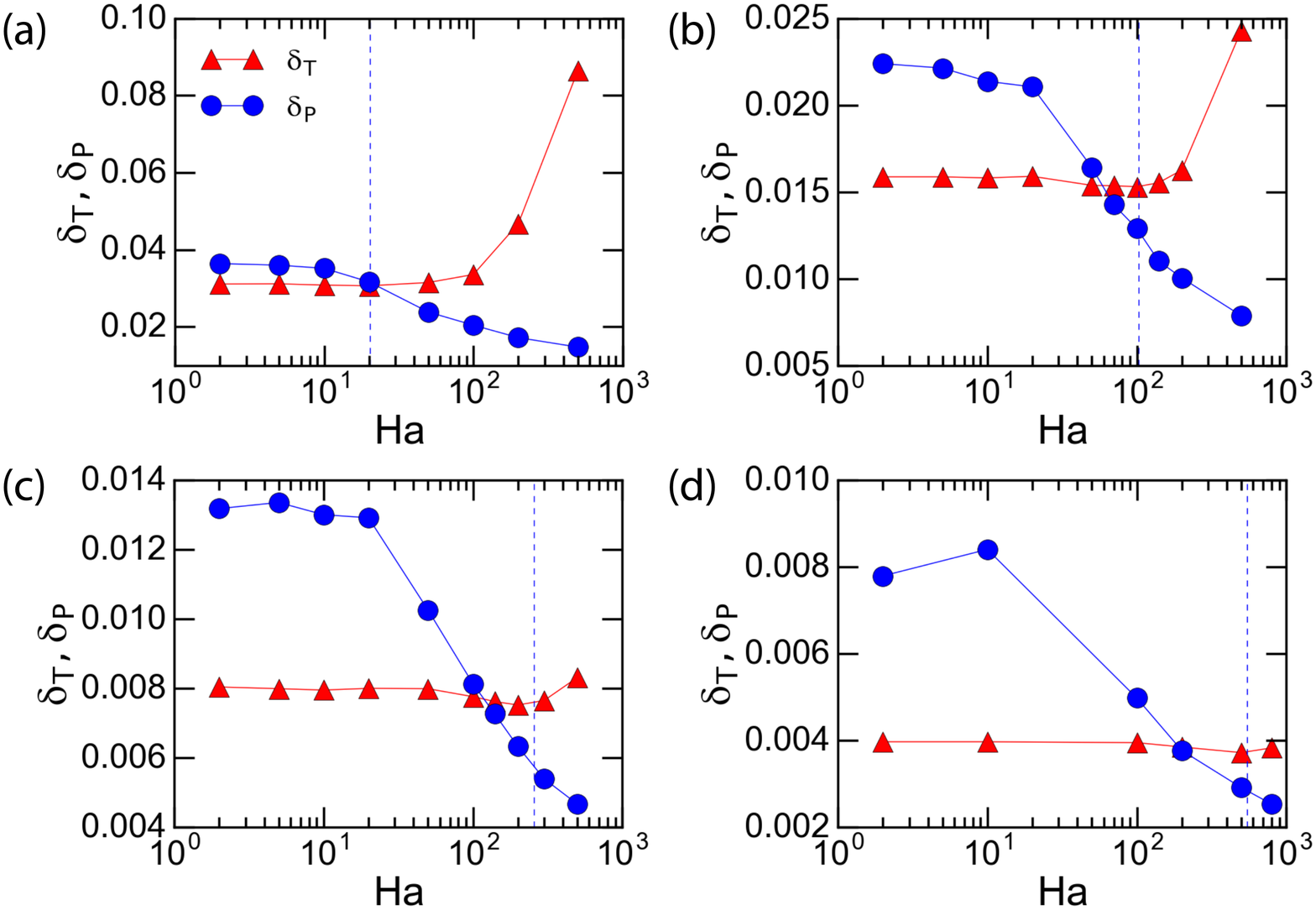}
\caption{Thermal $\delta_T$ and momentum (stress) $\delta_P$ boundary layer thicknesses as a function of the Hartmann number $Ha$.  (a) - (d) refers to the cases $Ra= 10^{7}, 10^{8}, 10^{9}, 10^{10}$ respectively.  The vertical dashed lines denote the positions of $Ha_{opt}$ where the estimated maximum enhancement occurs.}
\label{fig:BLthickness}
\end{figure}

Following the idea proposed by \cite{chong2017prl}, we  define the thickness of the stress (momentum) boundary layer as the peak position of the profile of the quantity  $(\partial_{x}u)^{2}+(\partial_{y}v)^{2}+(\partial_{z}w)^{2}$. The quantity is the square of the normal gradient of velocity summing over all components, which measures the overall magnitude of the normal stress (hereafter simply referred to as the stress); in \cite{chong2017prl} it is called the momentum boundary layer. Note that this definition is different from the traditionally defined velocity boundary layer, which will be discussed in Sec. 3.5 and which will be called the {\it viscous boundary layer}. For the temperature boundary layer, we use the so-called slope method based on the mean temperature profile (\citefullauthor{tilgner1993temperature} \citeyear{tilgner1993temperature}; \citefullauthor{lui1998pre} \citeyear{lui1998pre}). The linear region of the horizontal velocity profile is close to 0, we fit the data points between $0 < z < z_{T,linear}$ for evaluating the BL thickness. $z_{T,linear}$ is given value of $0.01, 0.005, 0.002$ and 0.001 for $Ra = 10^7,10^8,10^9$ and $10^{10}$ respectively. The error of $\delta_T$ is determined by the difference between this result and that when one more data point is included in the fiting. 

 
Figure \ref{fig:profile} plots the profiles of the temperature and of the stress for the case of $Ra = 10^{10}$ and $Ha = 10$ (the inset shows the peak of stress profile more clearly in log scale for the horizontal axis). From the profile we obtain respectively the thickness  $\delta_T$ of the thermal boundary layer and the thickness $\delta_P$ of the momentum boundary layer, as indicated by the blue dashed vertical lines. Note that in the inset figure plotted in semi-log scale, data points with $0\leq z \leq 0.5$ are shown as red circles and those with $0.5\leq z \leq 1$ are projected under transformation $1-z$ are shown as blue triangles. That the positions of the peaks from near the top and from near the bottom plates coincide with each other indicates a nearly perfect top-bottom symmetry about the middle height of the system.

Figure \ref{fig:BLthickness} plots $\delta_T$ and $\delta_P$ as functions of $Ha$ for the four $Ra$ values respectively. An overall feature seen from these plots is that for $Ha$ smaller than a certain value, both $\delta_T$ and $\delta_P$ do not change significantly with $Ha$ and that the thermal boundary layer remains nested inside the stress (momentum) layer. When $Ha$ becomes larger than $\sim O(10)$, the momentum layer starts to decrease rapidly, while the thermal boundary experiences an increase. The onset $Ha$ value roughly corresponds to that for the onset of $Nu$ enhancement and $Re$ decrease. With increasing magnetic field, the two boundary layers eventually cross over at certain value of $Ha$, which itself increases with $Ra$. This crossover point is very close to the Hartmann number $Ha_{opt}$ for the optimal heat transfer enhancement, which are indicated as dashed vertical lines in the respective figures. Similar feature is also observed for the optimal heat/salt transport enhancement in the systems of CRB, RRB and DDC \citep{chong2017prl}.

\begin{figure}
\centering
\includegraphics[width=\textwidth]{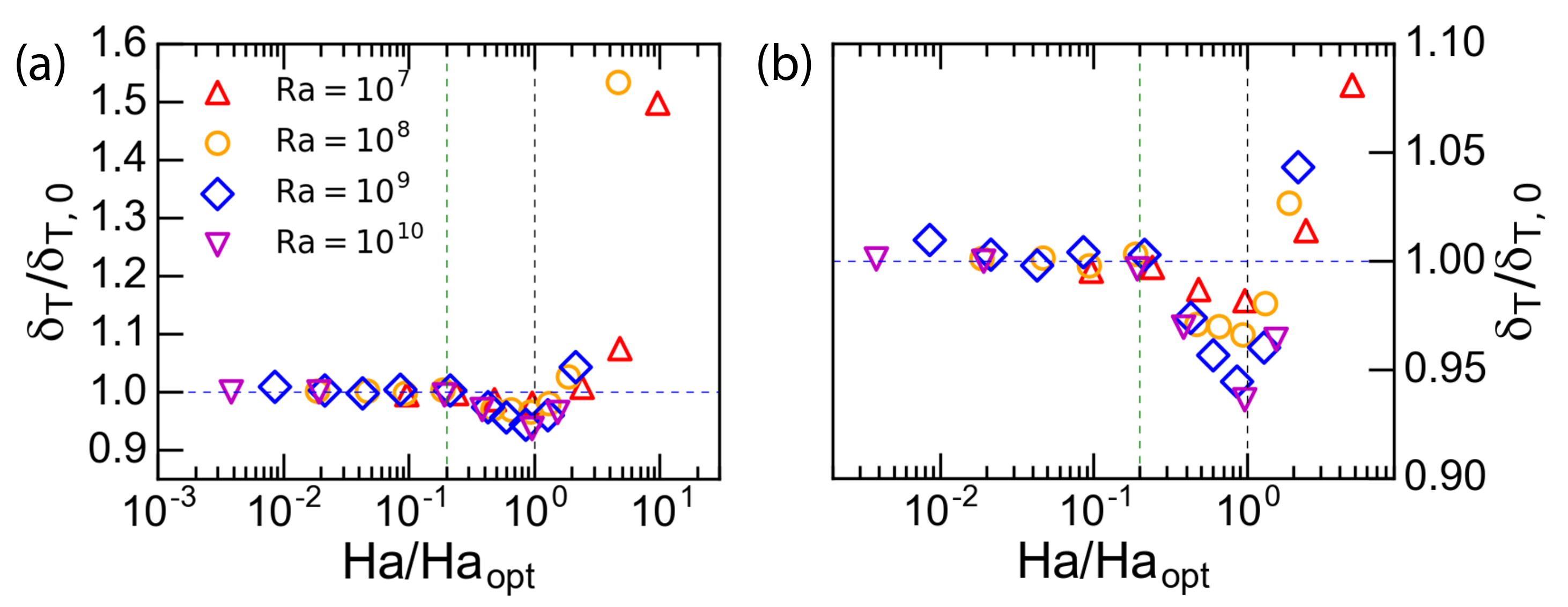}
\caption{(a) The thermal boundary layer $\delta_T$ normalised by the value at $Ha=0$ versus $Ha/Ha_{opt}$, for different values of $Ra$. (b) A magnified view of (a) to show more clearly the decrease of $\delta_T$, with the position of the minimum of thermal boundary layer corresponds to the maximum $Nu$ shown in figure 5(c). The green dash line indicates $Ha = 0.2Ha_{opt}$ for the onset of heat transfer enhancement and the black dash line indicates $Ha = Ha_{opt}$.}
\label{THBLHa}
\end{figure}

A well-known property in RB convection is that the thermal BL thickness should decrease when $Nu$ is increased.  This feature is not obvious in figure \ref{fig:BLthickness},  because of the scale of the plots. To show that $Nu$ enhancement is indeed accompanied by a decrease in the thermal BL, we show in figure \ref{THBLHa}(a)  the thermal BL thickness normalised by its value under zero magnetic field versus $Ha/Ha_{opt}$, and an enlarged portion in figure  \ref{THBLHa}(b). It is evident from the figure that $\delta_T$ indeed starts to decrease at about the same $Ha$ value ($\sim 0.2Ha_{opt}$) as $Nu$ starts to increase, reaching a minimum at the optimal Hartmann number and then starts to increase sharply, which corresponds to the sharp $Nu$ decrease. We remark that, as the Reynolds number has decreased sharply under the stabilising Lorentz force, the thermal boundary layer is not thinned by a stronger shear in this case. As shown by \cite{chong2015prl}, the thinning of  $\delta_T$ is a result of the increased plume coherency, so that the hot (cold) plumes are able to more efficiently cool (heat) the thermal boundary layer when reaching the opposite plate after traversing the bulk of the fluid within which the stabilising effect takes place. This is another example of what is termed plume-controlled regime in which the boundary layers are being controlled, via the modification of the bulk flow, rather than controlling \citep{chong2015prl}.

\begin{figure}
\centering
\includegraphics[width=\textwidth]{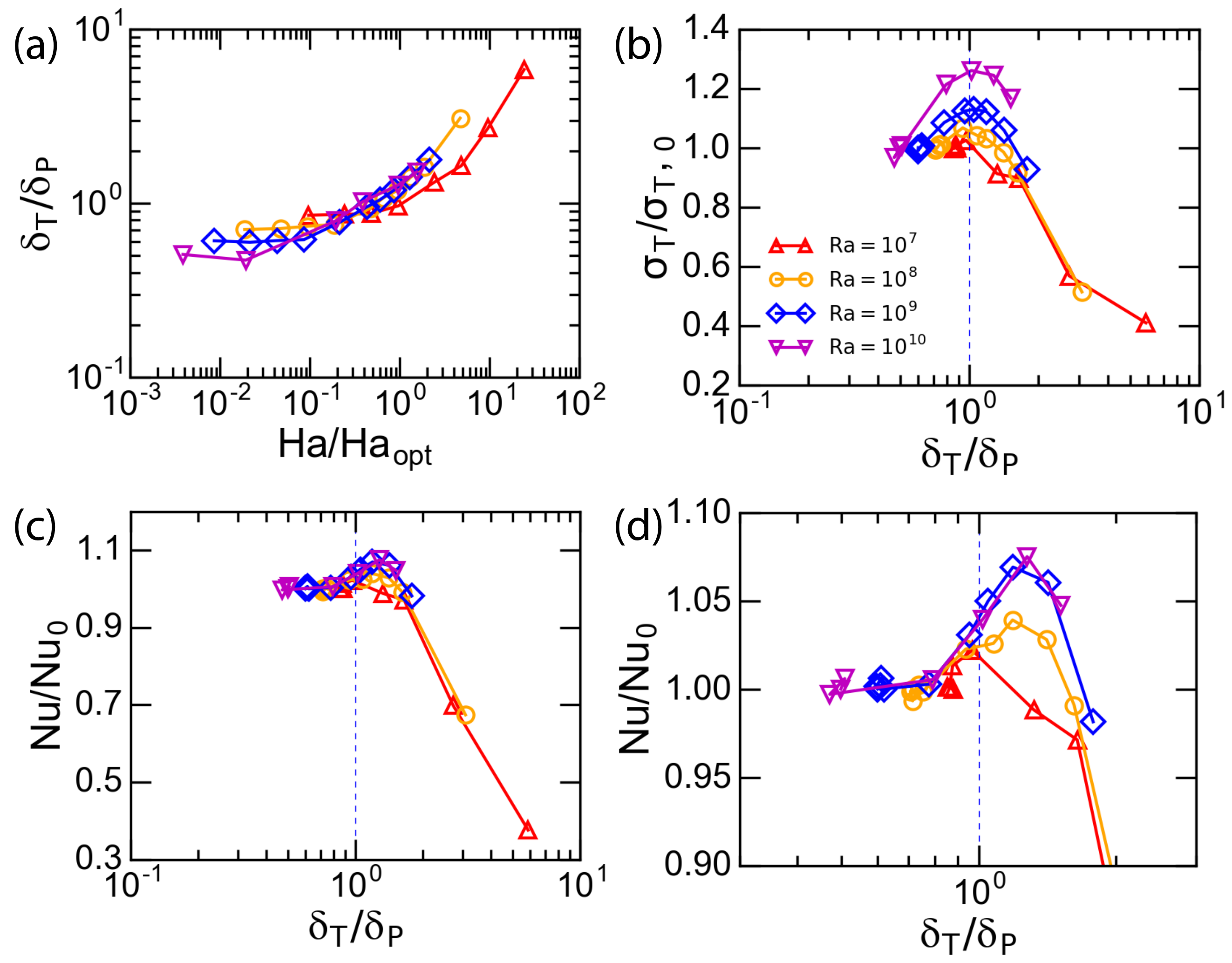}
\caption{Correlation of the boundary layer thickness ratio $\delta_T/\delta_P$ with various quantities. (a) $\delta_T/\delta_P$  vs $Ha/Ha_{opt}$. (b) Normalised temperature standard deviation vs BL thickness ratio. (c) Normalised $Nu$ vs. $\delta_T/\delta_P$. (d) A magnified view of (c).}
\label{fig:BLratio}
\end{figure}

To further examine the role played by boundary layer crossing in determining the optimal heat transfer enhancement, we plot in figure \ref{fig:BLratio} the boundary layer ratio $\delta_T/\delta_P$ against various quantities. Figure \ref{fig:BLratio}(a) plots $\delta_T/\delta_P$ versus $Ha/Ha_{opt}$, which shows that the optimal Hartmann corresponds to the situation when the momentum BL becomes thinner than the thermal BL; and this property holds for all the $Ra$ values explored. Figure \ref{fig:BLratio}(b) plots $\sigma_{T}/\sigma_{T,0}$ versus $\delta_T/\delta_P$,  where  $\sigma_{T}$ is the temperature standard deviation at the edge of the thermal BL and $\sigma_{T,0}$ is its value at $Ha = 0$. This figure shows that when the thicknesses of the two BLs become comparable to each other, temperature fluctuations are enhanced and become maximum at the point of BL crossing. This enables the optimal coupling of the strongest suction with the maximal temperature fluctuations. 
It is known that thermal plumes are generated by thermal BL instability and detachment. Therefore, the above optimal coupling would enhance thermal plume emission, which in turn enhances heat transfer.
Figures \ref{fig:BLratio}(c) and (d) (an enlarged part of (c)) plot $\delta_T/\delta_P$ versus the normalized $Nu/Nu_{0}$, again, the figures show that optimal heat transfer enhancement corresponds to when the two BLs cross each other, or shortly thereafter. These features are exactly the same as those exhibited by the other three systems of stabilising-destabilising turbulent flows, i.e. the CRB, RRB and DDC as shown by \cite{chong2017prl}. This provides a very convincing evidence that the present system of quasistatic magnetoconvection belongs to the same universality class as the other three systems.

\subsection{Scaling behaviour of the global Nusselt number}

In this section we will examine the scaling behaviour of the global heat transport in  
quasistatic magnetoconvection under a vertical external magnetic field.

\begin{figure}
\centering
\includegraphics[width=\textwidth]{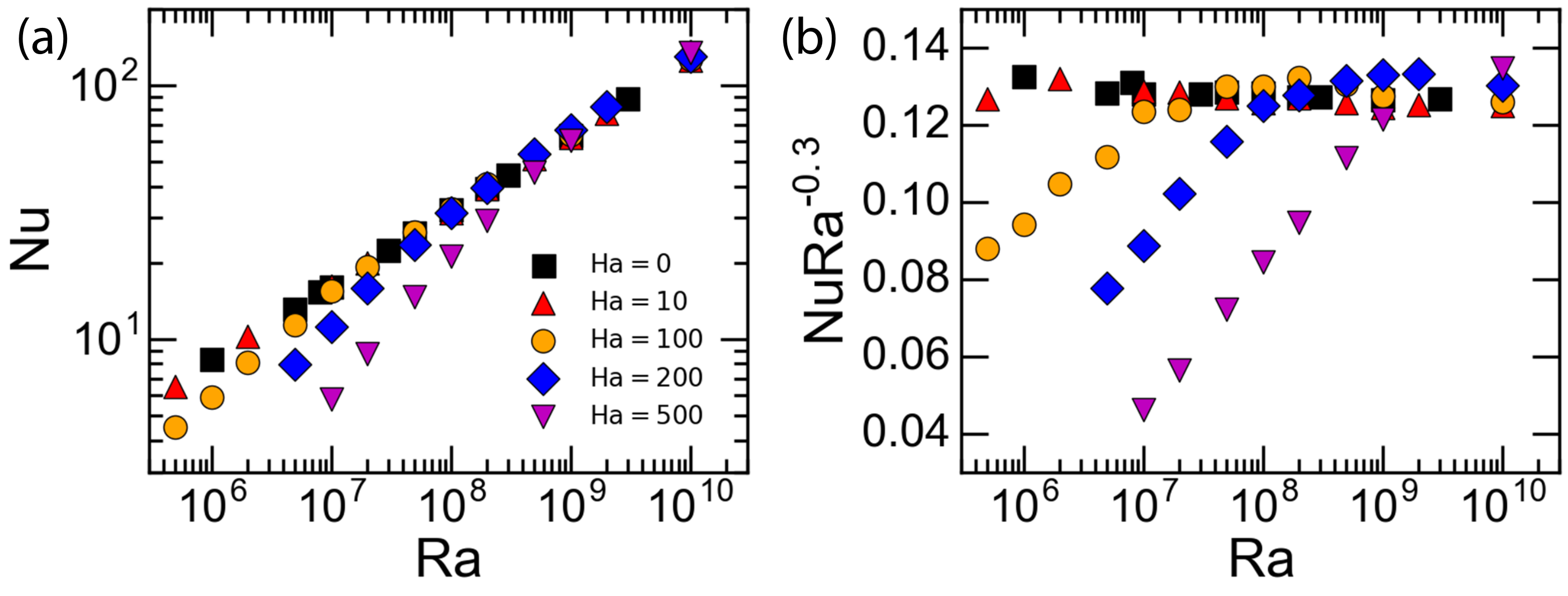}
\caption{$Nu$-$Ra$ plot for various values of $Ha$. (a) $Nu$ vs. $Ra$ for the cases of $Ha =  0, 10, 100, 200, 500$. The $Ha = 0$ ($Pr =4.38$) case is from \citefullauthor{kaczorowski2013jfm} \citeyear{kaczorowski2013jfm} and the data is used here as a baseline. (b) Compensate plot of $Nu$ with $Ra^{-0.3}$.}
\label{NuRa}
\end{figure}

Figure \ref{NuRa}(a) plots $Nu$ vs $Ra$ for the various values of $Ha$. For the case of $Ha =0$, we take the data set from \citefullauthor{kaczorowski2013jfm} (\citeyear{kaczorowski2013jfm}) and \citefullauthor{kaczorowski2014jfm} (\citeyear{kaczorowski2014jfm}) and use it as the baseline, which gives a scaling $Nu = 0.14Ra^{0.30}$. The range of parameter explored here spans from $Ra = 10^{6}$ to $10^{9}$. Within this range, it is seen that for $Ha =10$ the scaling is essentially the same as the baseline case. For $Ha \geq 100$, below certain value of $Ra$, the scaling start to deviate from the baseline. Figure 9(b) plots the $Nu$ data compensated by the scaling exponent from the baseline data. Here, one sees that for $Ha = 100$, the $Nu$-$Ra$ scaling follows the classical Rayleigh-B\'enard scaling for $Ra \geq 2 \times 10^{7}$. For $Ra \leq 2 \times 10^{7}$, a much steeper scaling of Nu $\sim Ra^{0.42}$ is observed. The $Ha = 200$ and $Ha = 500$ cases also exhibit a similar transition at $Ra = 2 \times 10^{8}$ and $2 \times 10^{9}$, respectively. For $Ha = 200$, the steeper scaling is $Nu \sim Ra^{0.50}$, while for $Ha = 500$, the scaling becomes $Nu \sim Ra^{0.58}$ for lower values of $Ra$. To better understand the observed scaling transition, we introduce a new parameter called the transition Rayleigh number, $Ra_{T}$, which, following Chong \& Xia for the case of severely-confined RB convection, can be defined by generalising the relationship between $Ha_{opt}$ and $Ra$ (see figure 5(d)) as

\begin{equation}
Ra_T=(\frac{Ha_{opt}}{0.018})^{1/0.45}
\end{equation}

\begin{figure}[tb]
    \centering
    \includegraphics[width=3.5in]{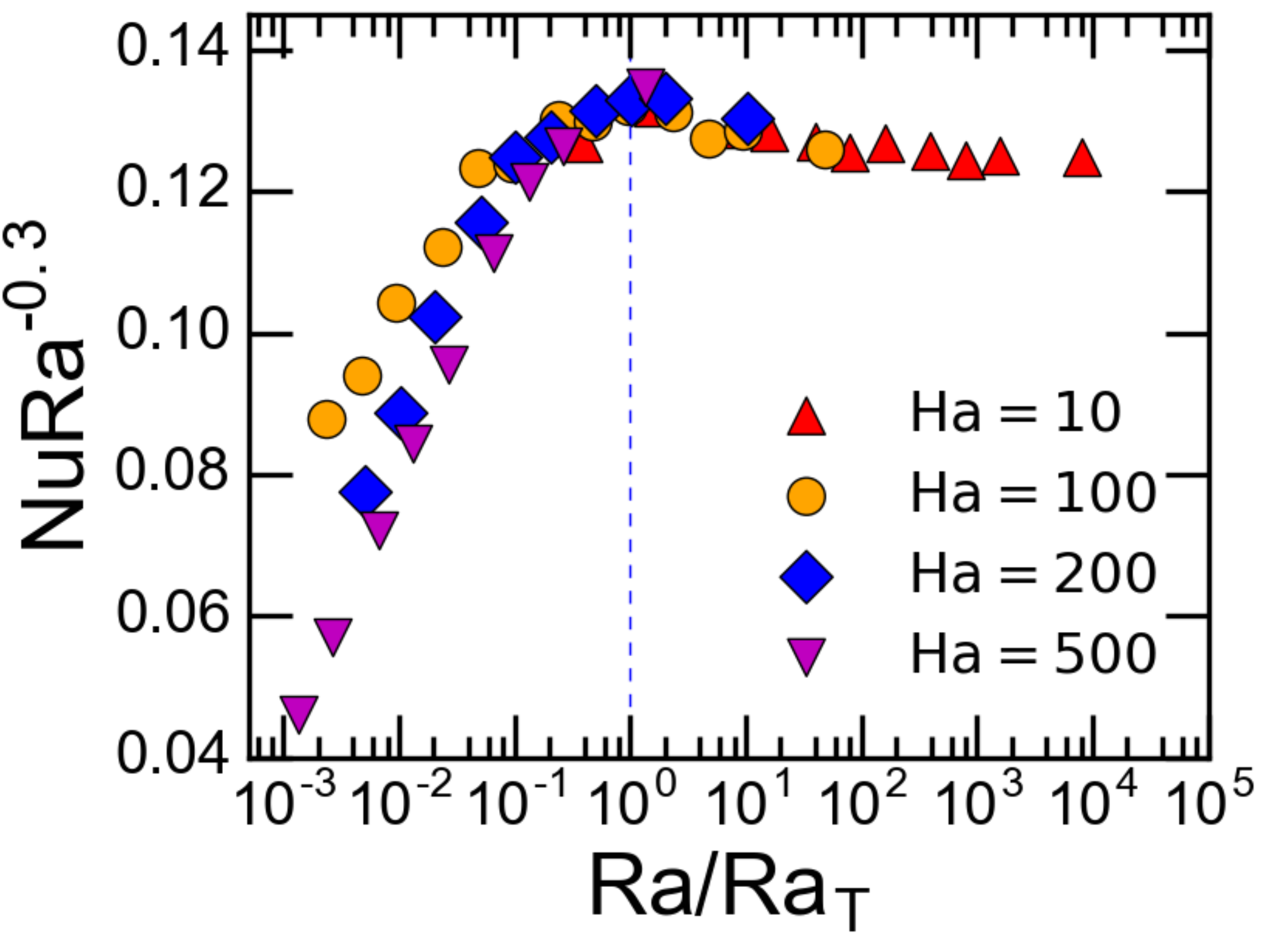}
    \caption{Compensate plot of $NuRa^{-0.3}$ with Ra normalized by $Ra_T$. Dash line indicates $Ra=Ra_T$.}
    \label{compensateplot}
\end{figure}

In figure \ref{compensateplot} we replot the data by the compensated $Nu$, but with $Ra$ normalised by $Ra_T$. The figure shows that the transitions for the different values of $Ha$ all occur approximately at $Ra = Ra_T$ and that the data all collapse together for $Ra > Ra_T$, suggesting that the $Ha$-dependent $Ra_T$ defined above may be used as a characteristic Rayleigh number for identifying the regime transition. 
As $Nu$ decreases sharply after reaching the optimal enhancement, this suggests that the steeper scaling for $Ra < Ra_T$ is related to the reduction of heat transport. Again, this feature is similar to those found in RRB and CRB (\citeauthor{king2009boundary} \citeyear{king2009boundary}; \citefullauthor{chong2016jfm} \citeyear{chong2016jfm}).

\begin{figure}[tb]
    \centering
    \includegraphics[width=3.0in]{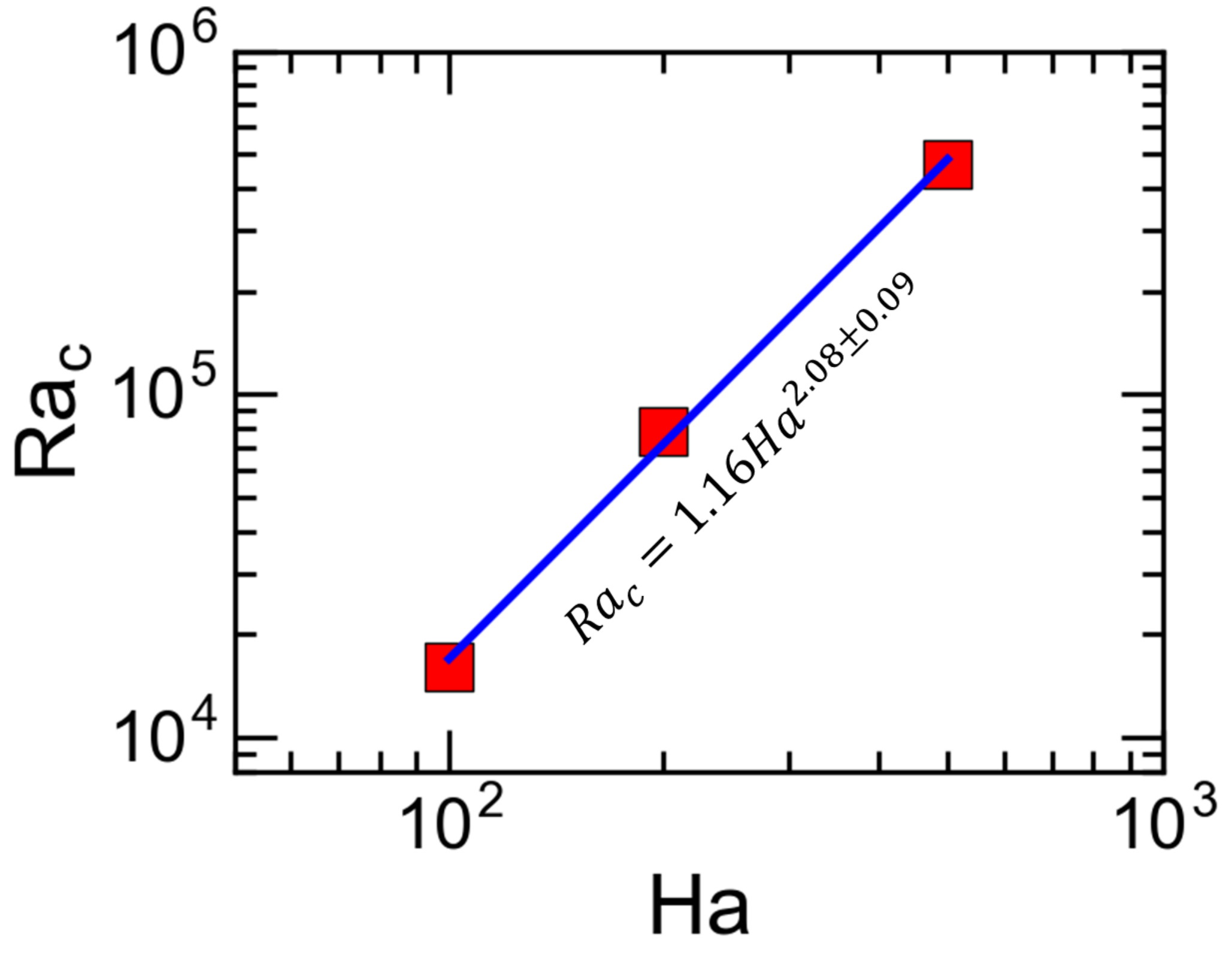}
    \caption{The onset Rayleigh number for convection, $Ra_{c}$ vs. $Ha$. A power law fit represented by the solid line gives $Ra_{c} =  1.16Ha^{2.08 \pm 0.09}$.}
    \label{Racrit}
\end{figure}

With the much steeper $Nu$-$Ra$ scaling and the suppressed heat transport for $Ra < Ra_T$, one would expect the onset value $Ra_c$ for convection will be dependent on $Ha$. To find out this, we perform an extrapolation taking the lowest 3 $Nu$ values for each $Ha$ to determine $Ra_{c}$. The estimated $Ra_{c}$ values are plotted in figure \ref{Racrit}. (For $Ha=10$, the scaling change is so small that we cannot confidently determine an $Ra_{c}$ for this case.)
A power-law fit to the data gives $Ra_{c} = 1.16Ha^{2.08 \pm 0.09}$. This value is in close agreement to the theoretical value of $Ra_{c} \sim Ha^{2}$ predicted by Chandrasekhar for magnetoconvection \citep{chandrasekhar2013hydrodynamic,aurnou2001experiments,de2017magnetoconvection}.


\subsection{The viscous boundary layer in magnetoconvection: Prandtl-Blasius-type vs Hartmann-type}

\begin{figure}[tb]
\centering
\includegraphics[width=\textwidth]{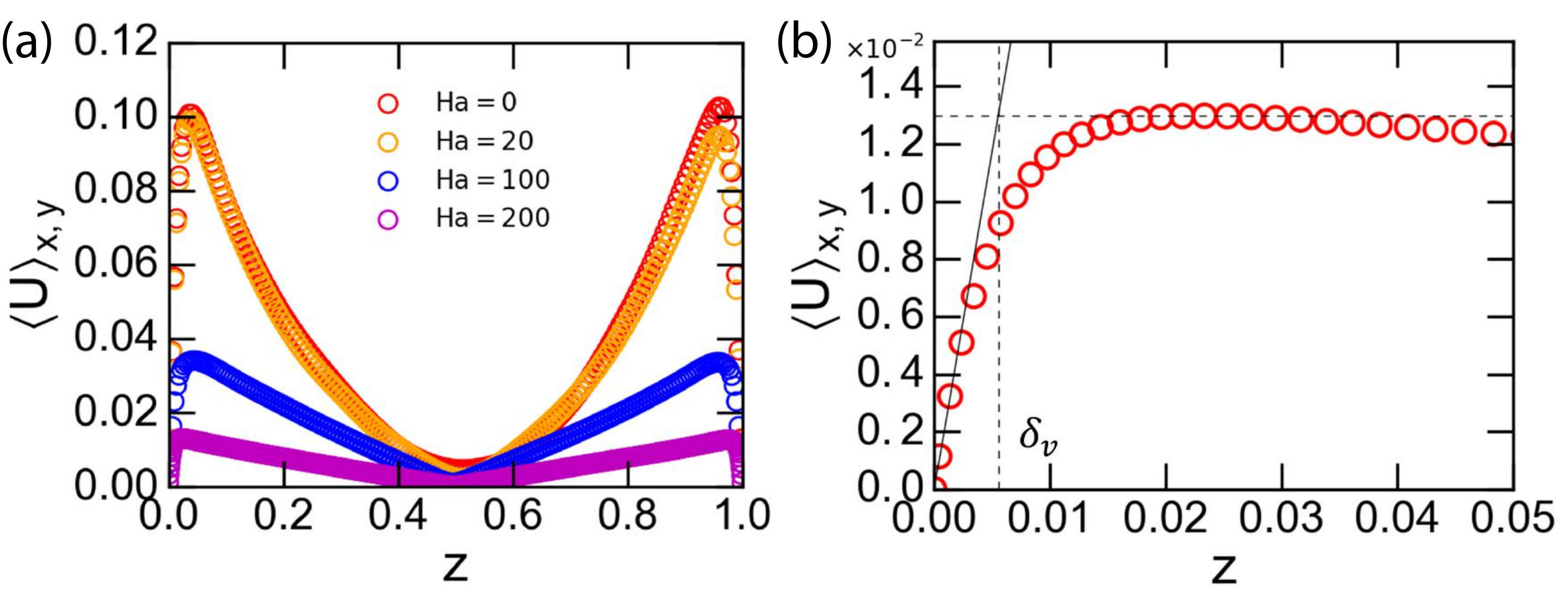}
\caption{(a) Mean horizontal velocity profile for different values of $Ha$ at $Ra= 10^9$. (b) An example of mean horizontal velocity profile close to the boundary layer ($Ra= 10^9$ and $Ha=200$). The lines define the thickness of viscous boundary layer $\delta_v$.}
\label{fig:momenBL}
\end{figure}

In this subsection, we examine the properties of the traditionally defined viscous boundary layer thickness $\delta_v$, in particular on the boundary layer transition from Prandtl-Blasius-type to Hartmann-type. Figure \ref{fig:momenBL}(a) plots examples of the horizontal velocity profile, which shows clearly the existence of well-defined peaks near both the top and bottom plates, with the maximum velocity decreases with increasing $Ha$. Also note that, as a closed system, the mean flow velocity decays to zero at cell centre. Figure \ref{fig:momenBL}(b) shows a magnified view of the profile near the bottom plate ($Ra= 10^9$ and $Ha=200$), where it is seen that the peak is rather broad under this resolution. We therefore adopt the so-called slope method to determine the BL thickness $\delta_v$, which is illustrated in the figure. Again, we fit the data points between $0 < z < z_{v,linear} = 0.005$ for evaluating the BL thickness. The error of $\delta_v$ is determined by the difference between this result and that when one more data point is included in the fitiing. For cases with large $Ra$ and large $Ha$, indicated by * in table 1, $z_{v,linear} = 0.002$ is used and the error is determined similarly.  

 With the BL thickness determined, we now examine its properties for various values of $Ra$ and $Ha$. Figure \ref{fig:General,delta_v_Ha,deltavRa}(a) plots in log-log scale $\delta_{v}$ vs. $Ra$ for various values of $Ha$. If one starts with large $Ha$, then it is seen that $\delta_{v}$ appears to be insensitive to changes in $Ra$. As $Ha$ decreases, $\delta_{v}$ converges gradually to its $Ha=0$ value. For example, for $Ha= 0$, $2$, and $10$, the values of $\delta_{v}$ are almost the same, indicating that the viscous boundary layer is not appreciably perturbed under the corresponding magnetic field. Furthermore, a simple power law relationship between $\delta_{v}$ and $Ra$ is evident for these low $Ha$ data.  In figure \ref{fig:General,delta_v_Ha,deltavRa}(b) we plot $\delta_{v}$ vs $Ha$ for various values of $Ra$. It is seen that for weak magnetic field, $\delta_{v}$ is independent of $Ha$ and its magnitude depends on $Ra$ only. For large values of $Ha$, on the other hand,  $\delta_{v}$ for different $Ra$ appears to converge to a value independent of $Ra$ and depends only on $Ha$. Therefore, it is clear that $\delta_{v}$ exhibits two types of asymptotic behaviour, depending on the relative magnitude of $Ra$ (or $Re$) and $Ha$, or the relative strength of driving over stabilising forces.

\begin{figure}[tb]
\centering
\includegraphics[width=0.64\textwidth]{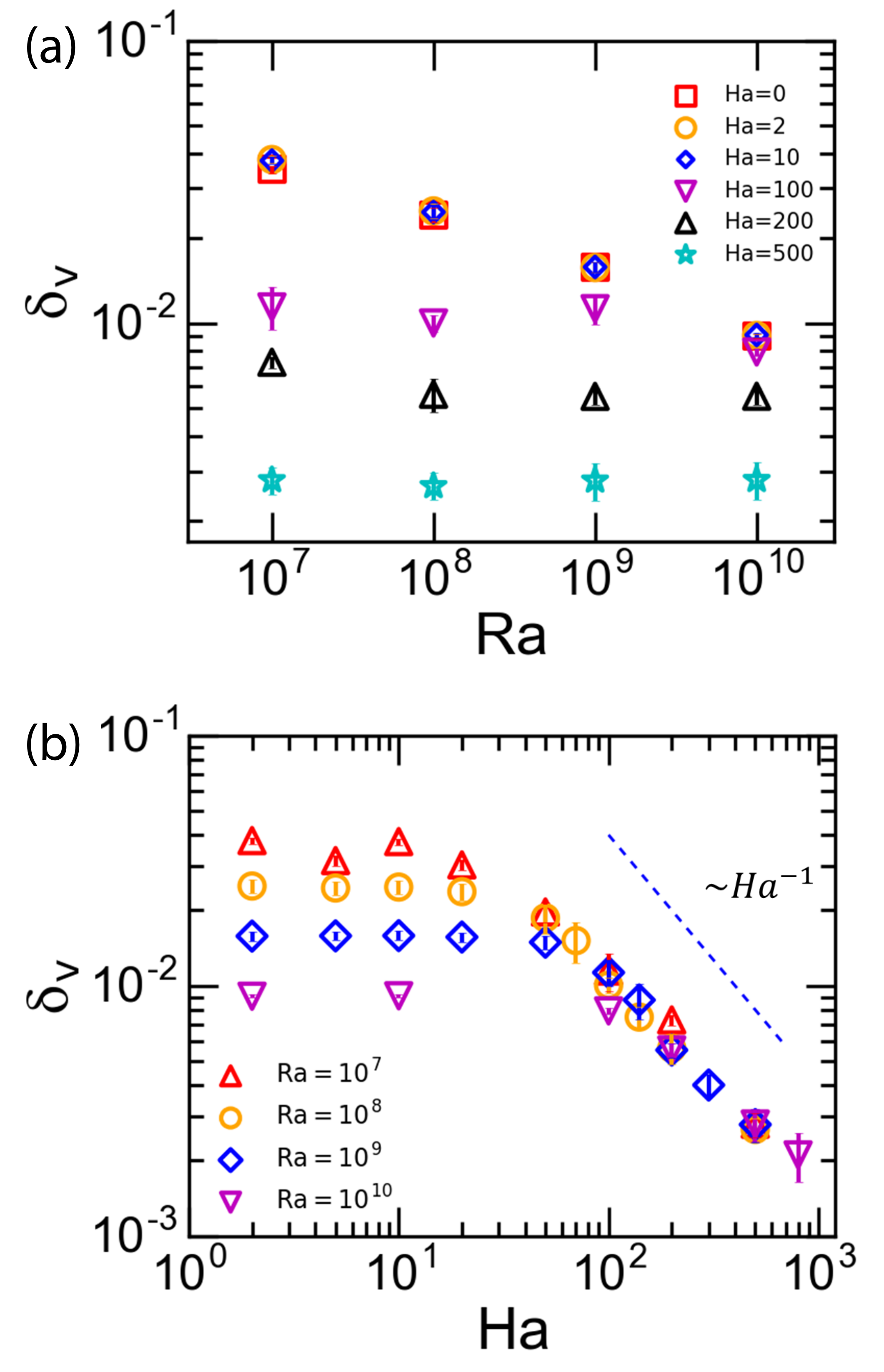}
\caption{Viscous boundary layer $\delta_v$ dependence on $Ra$ and $Ha$. (a) $\delta_v$ vs $Ra$ for various values of $Ha$. (b) $\delta_v$ vs. $Ha$ for various values of $Ra$; the dashed line indicates an $Ha^{-1}$ scaling.}
\label{fig:General,delta_v_Ha,deltavRa}
\end{figure}

\begin{figure}[tb]\centering
\includegraphics[width=\textwidth]{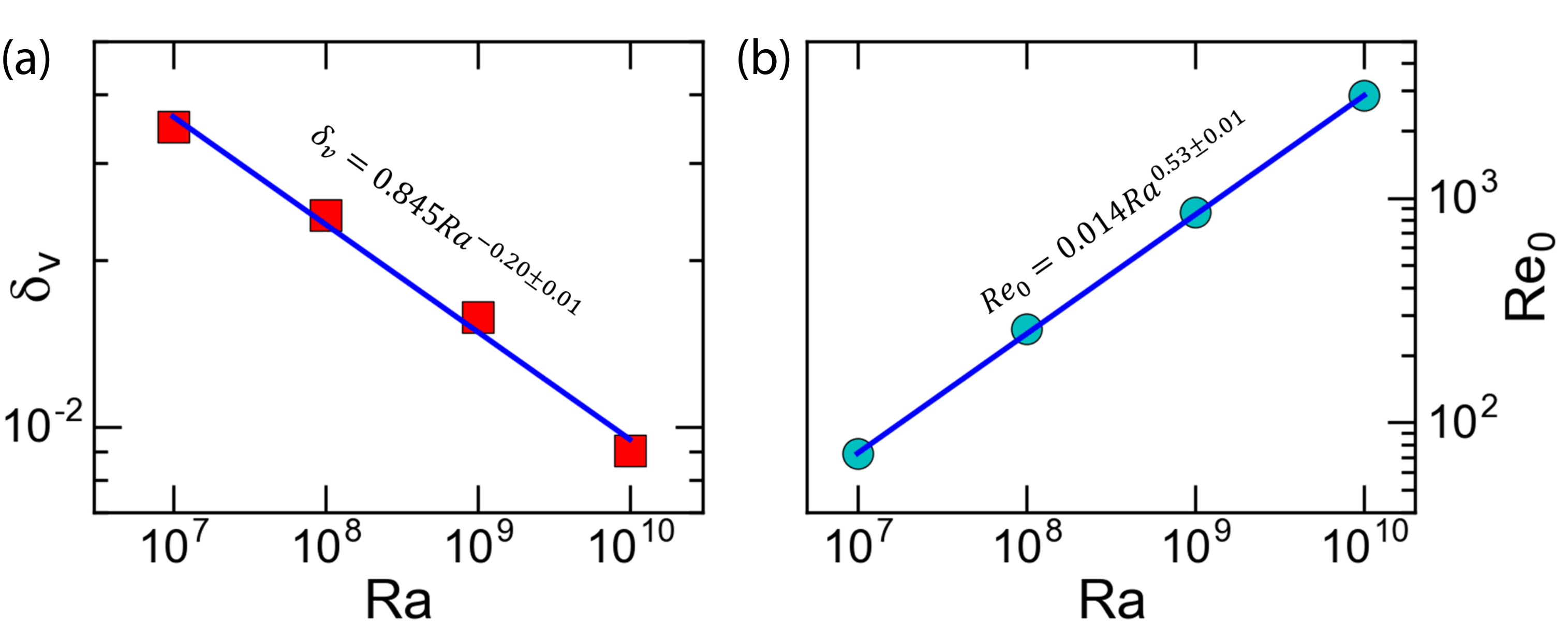}
\caption{Scaling properties of the viscous boundary layer and the Reynolds under zero magnetic field. (a ) The dependence of $\delta_v$ on $Ra$ at $Ha=0$; a power law fit represented by the solid line yields $\delta_v = 0.845Ra^{-0.20\pm 0.01}$. (b) The Reynolds number $Re_0$ at $Ha=0$ versus $Ra$, where a power law fit yields $Re_0 = 0.014Ra^{0.53\pm 0.01}$.}
\label{fig:deltav_Ra_Ha0}
\end{figure}

To gain insight into the dependence of $\delta_{v}$ on $Re$ and $Ha$, we consider two asymptotic limits, i.e. $Re \gg Ha$ and $Re \ll Ha$.  In the classical regime of turbulent thermal convection, i.e. the BL remains laminar, and in the absence of magnetic field (corresponding to $Re \gg Ha$), one can obtain the Prandtl-Blasius-type BL by balancing the inertial term and the viscous term in the momentum equation, which gives:

\begin{equation}
\delta_v =\frac{H}{\sqrt{Re}}
\end{equation}

In the limit of $Re \ll Ha$,  the Lorentz force becomes dominant and the viscous term balances the Lorentz term and one obtains the so-called Hartmann boundary layer  (\citefullauthor{zurner2016heat} \citeyear{zurner2016heat}):

\begin{equation}
\delta_v = \frac{H}{Ha}
\end{equation}

For the intermediate cases, we assume that both the inertial and Lorentz terms are important in the momentum equation and it is the inertial term that balances both the viscous and the Lorentz terms, i.e.
 
\begin{equation}
(\vec{u} \cdot \nabla)\vec{u} \sim \nu \nabla^2 \vec{u} + \frac{1}{\rho}(\vec{J} \times \vec{B}) 
\label{forcebalance}
\end{equation}

 Using dimensional analysis, we can write (\ref{forcebalance}) as
\begin{equation}
\frac{U^2}{H} \sim \frac{\nu U}{\delta^2_v} - \frac{\nu UHa^2}{H}
\end{equation}
which gives
\begin{equation}
\label{theoreticalscaling}
\delta_v \sim \frac{H}{\sqrt{Re + Ha^2}}
\end{equation}

As the obtained $\delta_{v}$ are in terms of the two control parameters $Ra$ and $Ha$ of the study, we need to first determine its $Re$ dependence.  To obtain the asymptotic behaviour of $\delta_{v}$ in the limit of $Re \gg Ha$, we fit a power law to its values at $Ha = 0$, which is shown in figure \ref{fig:deltav_Ra_Ha0}(a). The obtained power law exponent of $-0.20$ agrees excellently with previous measured values (see, for example, \citefullauthor{xin1996measured} \citeyear{xin1996measured}; \citefullauthor{wei2013viscous} \citeyear{wei2013viscous}). For the $Re$-$Ra$ relationship, we show in figure \ref{fig:deltav_Ra_Ha0}(b) a plot of the zero-$Ha$ Reynolds number $Re_0$ versus $Ra$; a power law fit gives $Re_0=0.014Ra^{0.53\pm0.01}$.
Although the viscous BL in RB convection remains largely laminar in the so-called classical regime,  it is rare that the exact Prandtl-Blasius (PB) scaling $\delta_v \sim Re^{\beta}$ with $\beta =1/2$ is observed (see, for example, \citefullauthor{xin1996measured} \citeyear{xin1996measured}; \citefullauthor{sun2008experimental} \citeyear{sun2008experimental}; \citefullauthor{wei2013viscous} \citeyear{wei2013viscous}). Part of the reason for the discrepancy is that the PB boundary layer theory is two-dimensional and most of the experimental and numerical studies are inherently three-dimensional. In general, the exponent $\beta$ appears to depend on the geometry and also the location (i.e. sidewall or horizontal plate) of the measurement and its value varies from $-0.32$ to $-0.5$ \citep{wei2013viscous}. In the present work we find $\delta_{v} \sim Ra^{-0.20}$ and $Re \sim Ra ^{0.53}$ for the asymptotic case of $Ha \rightarrow 0$, which implies $\delta_{v} \sim Re^{-0.38}$ in the $Ha \rightarrow 0$ limit. Therefore, we can write 

\begin{figure}[tb]
        \centering
    \setlength{\unitlength}{\textwidth}
    \includegraphics[width=1.0\unitlength]{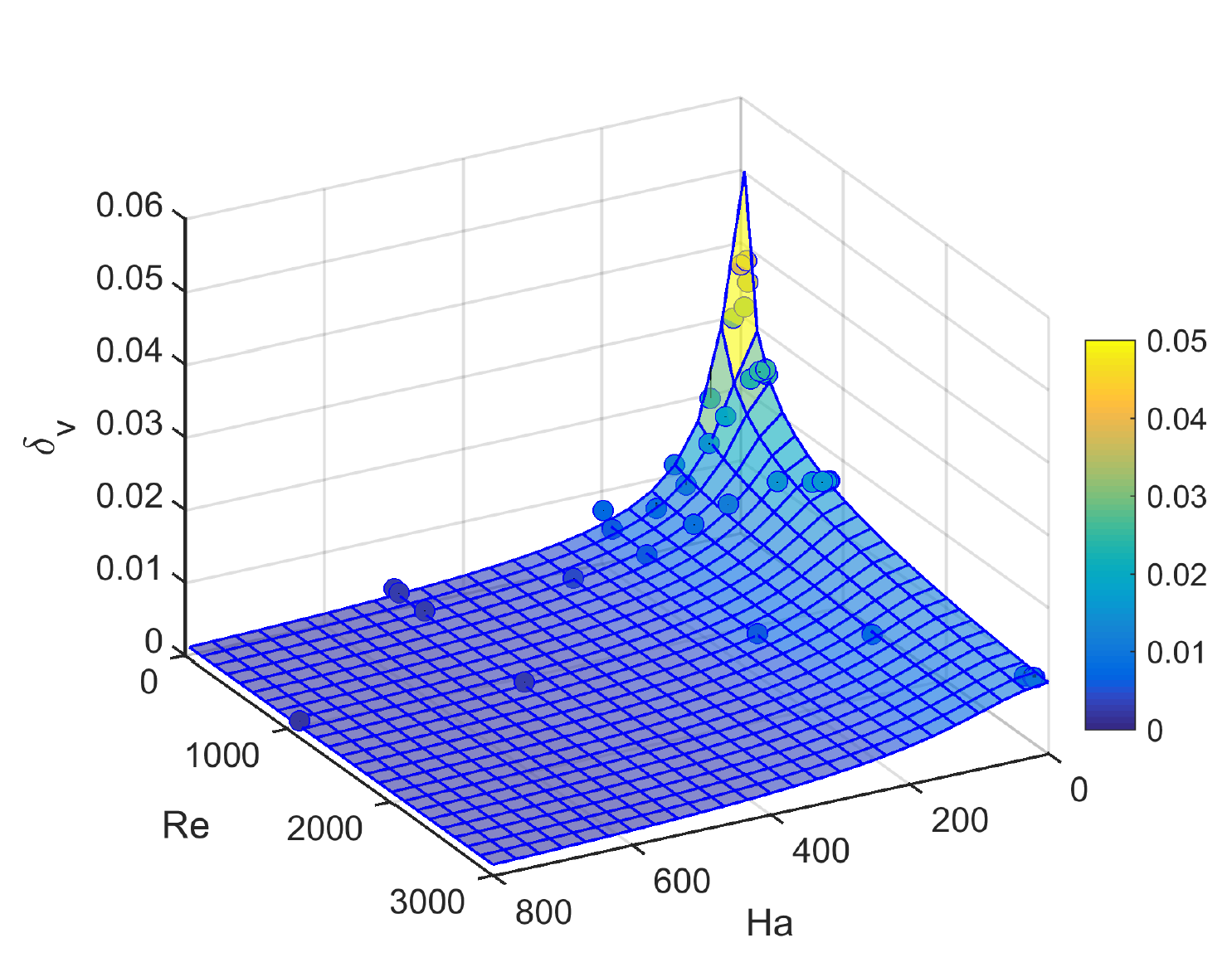}
\caption{Three dimensional surface plot of the dependence of the viscous boundary layer on both $Re$ and $Ha$. Circles represent the numerically obtained $\delta_v$  and the surface  represents equation $\delta_v= H/{\sqrt{24Re^{0.76} + 0.67Ha^2}}$.}
\label{fig:BLfit}
\end{figure}
\begin{figure}[tb]
        \centering
    \setlength{\unitlength}{\textwidth}
    \includegraphics[width=1.0\unitlength]{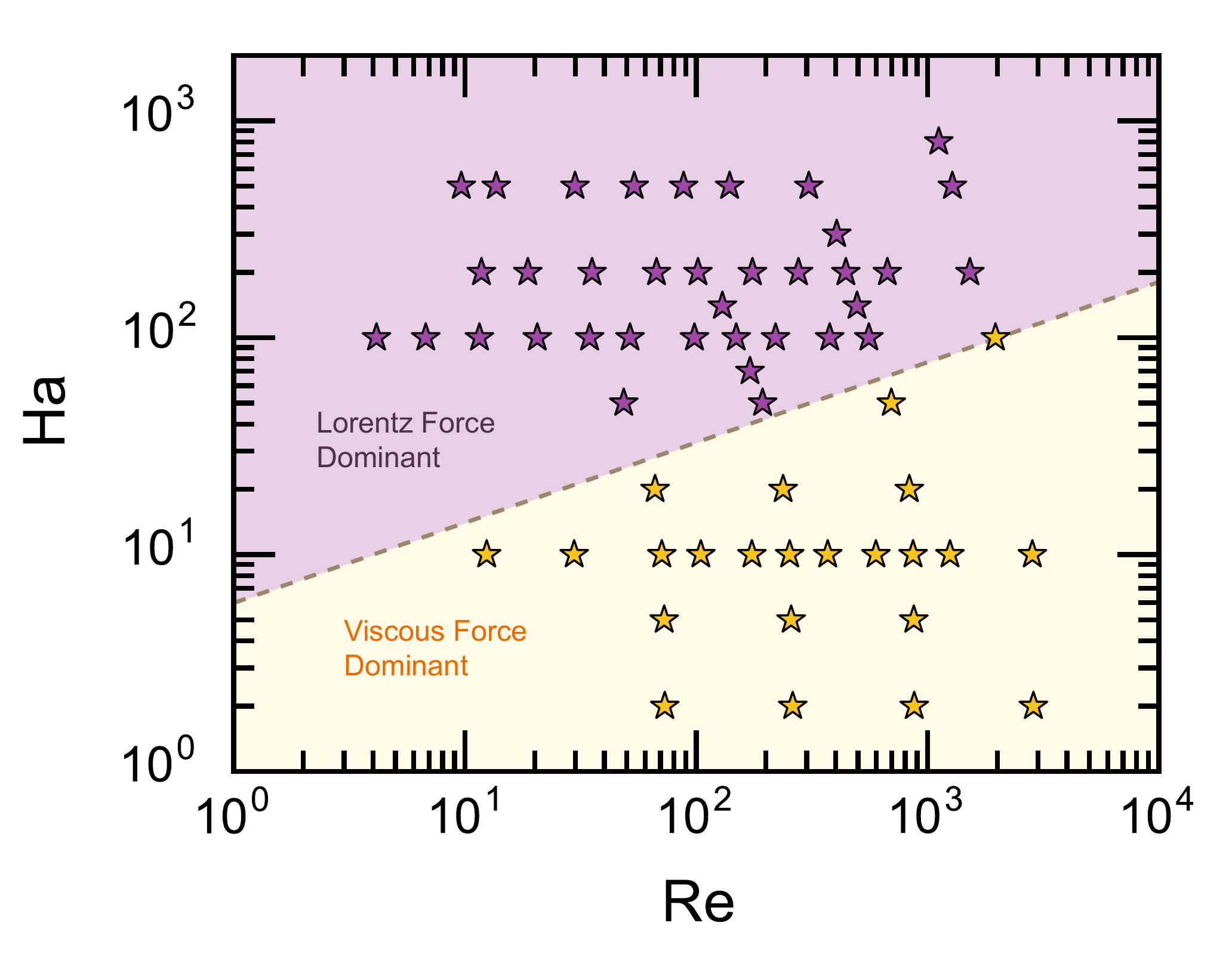}
\caption{Boundary layer phase diagram. The cream-coloured area denotes the region where the boundary layer is dominated by the viscous force, and the purple-coloured area denotes the region dominated by the Lorentz force. The dash line indicates where the contributions by the two forces to the boundary layer equal to each other, i.e. $24Re^{0.76} = 0.67Ha^2$. The stars represent the parameters for which simulations were made.}
\label{fig:dominant}
\end{figure}


\begin{equation}
\label{scaling}
\delta_v = \frac{H}{\sqrt{c_1Re^{0.76} + c_2Ha^2}}
\end{equation}
which would yield the numerically determined Prandtl-Blasius-like scaling and the Hartmann-like scaling in the respective limits of weak and strong magnetic fields, but also gives the $Re$- and $Ha$-dependence of $\delta_v$ for the intermediate cases. In the above the coefficients $c_1$ and $c_2$ account for the relative contributions of the inertial and Lorentz forces. By fitting the above equation to $\delta_v$ for the various values of $Re$ and $Ha$, we obtain $c_1 = 24$ and $c_2 = 0.67$. The fitting result is  shown in figure \ref{fig:BLfit}, where the surface represents equation (\ref{scaling}) with the fitted parameters; and the circles are the numerically obtained $\delta_v$. It is seen that almost all data points fall onto the curved surface. To see the $Re$-dominant and $Ha$-dominant regions more clearly, we plot in figure \ref{fig:dominant} a phase diagram in which the cream-coloured area denotes the region where the boundary layer is controlled by the viscous force and the purple-coloured area denotes the Lorentz force controlled region. The two regions are separated by the dash line, which is determined by setting $24Re^{0.76} = 0.67Ha^2$, i.e. the contributions by the two forces to the boundary layer equal to each other. Thus, the behaviour of the viscous boundary layer in quasistatic magnetoconvection may be understood as fallows: When $Ha$ is very small, the viscous boundary layer is dominated by the inertial term and it behaves as a Prandtl-Blasius-type BL which depends only on $Re$;  as $Ha$ increases, the Lorentz force becomes dominant and the BL becomes Hartmann-type layer and becomes independent of $Re$. In the intermediate range, the BL thickness may be represented by equation (\ref{scaling}).

\section{Conclusion}
We have made a numerical study of quasistatic magnetoconvection. Two sets of simulations were made. In the first one, the Hartmann number $Ha$ varied from 0 to 800 for each value of the Rayleigh number $Ra = 10^7,10^8,10^9$ and $10^{10}$. In the second set of data, $Ra$ varied from $10^5$ to $10^{10}$ for each value of $Ha = 10, 100, 200$, and 500. Our results show that as the strength of the magnetic field, represented by the Hartmann number $Ha$, is increased above certain threshold value, the flow strength as represented by the Reynolds number starts to decrease monotonically, which is understood as the suppression of the flow by the Lorentz force. The heat transport efficiency, as characterised by the Nusselt number, on the other hand, behaves differently. When $Ha$ is above the threshold value, $Nu$ first increases, representing an enhancement. 

With increasing $Ha$, $Nu$ reaches a maximum value for a certain optimal $Ha_{opt}$ that is $Ra-$dependent. The enhanced heat transport may be understood as a result of the increased coherency of the thermal plumes, which are elementary heat carriers of the system. As $Ha$ increases beyond $Ha_{opt}$, $Nu$ starts to decrease sharply, indicating that the effect of the suppression of the flow by the Lorentz force has now overtaken the benefit brought about by the increased plume coherency. To our knowledge this is the first time that a heat transfer enhancement by the stabilising Lorentz force in quasistatic magnetoconvection has been observed. We further found that the optimal enhancement may be understood in terms of the crossing between the thermal and the momentum (stress) boundary layers and the fact that temperature fluctuations are maximum near the position where the BLs cross, which suggests that the optimal enhancement of $Nu$ is related to the increased thermal plume emissions.
These findings demonstrate that the heat transport enhancement in the quasistatic magnetoconvection system belongs to the same universality class of stabilising$-$destabilising ($S$-$D$) turbulent flows as the systems of confined Rayleigh-B\'enard (CRB), rotating Rayleigh-B\'enard (RRB) and double-diffusive convection (DDC). This is further supported by the findings that the heat transport, boundary layer ratio and the temperature fluctuations in magnetoconvection at the boundary layer crossing point are similar to the other three cases. These four systems belong to the same universality class raises an interesting possibility that one or some of them may be used as a proxy for studying certain features of the other systems within the context of $S$-$D$ flows.

Based on the second set of simulations, a transition in the $Nu$-$Ra$ scaling is observed, such that below a transitional $Ra_{T}$, the $Nu$ is suppressed relative to its $Ha=0$ value, and its scaling with Ra becomes steeper. Moreover, the $Ha-$dependent $Ra_{T}$ is found to be a proper quantity to characerise the $Nu$ transition from classical to steeper scaling. When $Ra$ is normalised by $Ra_T$, it is found that the $Nu$-$Ra$ plot collapse into a general trend above $Ra_T$.  Since the transition is found near $Ra = Ra_T$, it is believed that the change of scaling behavior is related with the beginning of sharp drop of heat transport. It is also closely related to the intrinsic properties of boundary layers as the scaling transition occurs close to BL crossing as well.  

A second type of boundary layer-crossing is also observed in this work. In one limit ($Re \gg Ha$), we find that the viscous boundary ($\delta_v$) exhibits a Prandtl-Blasius-type scaling with the Reynolds number (or the Rayleigh number) and is independent of $Ha$. In the limit of $Re \ll Ha$, $\delta_v$ exhibits an approximate $\sim Ha^{-1}$ dependence, which has been predicted for a Hartmann boundary layer. Assuming the inertial term in the momentum equation is balanced by both the viscous and Lorentz terms, we derived an expression $\delta_v = H/\sqrt{c_1Re^{0.72} + c_2Ha^2}$ for all values of $Re$ and $Ha$, which fits the obtained viscous boundary layer well. Therefore, we can understand the viscous boundary layer behaviour as follows: For low $Ha$ and sufficiently large $Re$, the boundary layer remains laminar, i.e. Prandtl-Blasius-type. For moderate value of $Re$ and sufficiently large Ha, where the Lorentz force becomes dominant,  $\delta_v$ becomes Hartmann boundary layer. This could provide insight to the study of the boundary layer behaviour for other RB systems such as RRB where the viscous boundary layer is also governed by the force balance between inertial force, viscous force and Coriolis force in the momentum equation.

\section{Acknowledgements}
This work was supported by the Research Grants Council (RGC) of HKSAR (No. CUHK14301115 and CUHK14302317) and a NSFC/RGC Joint Research Project (Ref. NCUHK437/15).  

\clearpage

\appendix
\section{}\label{app}
\begin{table}[H]
\centering
\label{simulation table1}
\resizebox{0.7\textwidth}{!}{\begin{tabular}{lllllll}
\hline
$Ra$	&$Ha$	& &  $Nu$	&	$Re$	&   $\delta_v$ 			&$\delta_T$ \\
		&		& &			&			&	$(\times 10^{-2})$	&$(\times 10^{-3})$ \\
\hline 

1$\times 10^7$	&	0	& &	15.99$\pm$0.05	&	72.99	&	3.53$\pm$0.11&	31.22$\pm$0.44	\\
				&	2	& &	16.01$\pm$0.04	&	73.00	&	3.82$\pm$0.11&	31.07$\pm$0.45	\\
				&	5	& &	16.01$\pm$0.05	&	72.60	&	3.19$\pm$0.14&	31.13$\pm$0.44	\\
				&	10	& &	16.20$\pm$0.05	&	70.88	&	3.79$\pm$0.11&	30.80$\pm$0.45	\\
				&	20	& &	16.34$\pm$0.05	&	66.29	&	3.07$\pm$0.13&	30.61$\pm$0.45	\\
				&	50	& &	15.80$\pm$0.04	&	48.53	&	2.01$\pm$0.18&	31.47$\pm$0.41	\\
				&	100	& &	15.53$\pm$0.06	&	34.54	&	1.19$\pm$0.20&	33.55$\pm$0.44	\\
				&	200	&*&	11.16$\pm$0.03	&	18.69	&	0.75$\pm$0.04&	46.76$\pm$0.08	\\
				&	500	&*&	6.01$\pm$0.07	&	9.63	&	0.30$\pm$0.03&	86.38$\pm$0.09	\\
1$\times 10^8$	&	0	& &	31.42$\pm$0.26	&	261.28	&	2.44$\pm$0.15&	15.84$\pm$0.19	\\
				&	2	& &	31.38$\pm$0.26	&	260.93	&	2.52$\pm$0.15&	15.86$\pm$0.19	\\
				&	5	& &	31.21$\pm$0.25	&	256.88	&	2.48$\pm$0.15&	15.86$\pm$0.19	\\
				&	10	& &	31.49$\pm$0.24	&	253.19	&	2.50$\pm$0.16&	15.80$\pm$0.19	\\
				&	20	& &	31.38$\pm$0.25	&	237.14	&	2.40$\pm$0.18&	15.89$\pm$0.18	\\
				&	50	& &	32.13$\pm$0.24	&	193.31	&	1.90$\pm$0.25&	15.37$\pm$0.18	\\
				&	70	& &	32.23$\pm$0.23	&	170.53	&	1.56$\pm$0.28&	15.35$\pm$0.18	\\
				&	100	&*&	32.65$\pm$0.12	&	148.67	&	1.02$\pm$0.07&	15.29$\pm$0.04	\\
				&	140	&*&	32.30$\pm$0.12	&	129.57	&	0.77$\pm$0.07&	15.52$\pm$0.05	\\	
				&	200	&*&	31.13$\pm$0.11	&	101.60	&	0.58$\pm$0.08&	16.25$\pm$0.05	\\
				&	500	&*&	21.17$\pm$0.01	&	53.87	&	0.28$\pm$0.03&	24.29$\pm$0.03	\\
1$\times 10^9$	&	0	& &	61.94$\pm$0.31	&	868.04	&	1.60$\pm$0.06&	7.96$\pm$0.09	\\
				&	2	& &	62.06$\pm$0.41	&	874.89	&	1.60$\pm$0.07&	8.03$\pm$0.09	\\
				&	5	& &	62.04$\pm$0.33	&	871.07	&	1.60$\pm$0.06&	7.98$\pm$0.09	\\
				&	10	& &	62.32$\pm$0.33	&	864.49	&	1.60$\pm$0.07&	7.94$\pm$0.09	\\
				&	20	& &	61.94$\pm$0.36	&	833.11	&	1.58$\pm$0.07&	7.99$\pm$0.09	\\
				&	50	& &	62.11$\pm$0.27	&	696.42	&	1.52$\pm$0.09&	7.98$\pm$0.08	\\
				&	100	& &	63.84$\pm$0.34	&	556.87	&	1.17$\pm$0.13&	7.75$\pm$0.08	\\
				&	140	& &	65.03$\pm$0.30	&	495.02	&	0.92$\pm$0.14&	7.61$\pm$0.09	\\
				&	200	&*&	66.22$\pm$0.16	&	442.93	&	0.57$\pm$0.04&	7.51$\pm$0.02	\\
				&	300	&*&	65.68$\pm$0.22	&	404.19	&	0.42$\pm$0.04&	7.64$\pm$0.02	\\
				&	500	&*&	60.79$\pm$0.12	&	306.26	&	0.29$\pm$0.04&	8.30$\pm$0.02	\\
1$\times 10^{10}$	&	0	&*&	125.28$\pm$0.37	&	2870.52	&	0.91$\pm$0.01&	3.96$\pm$0.01	\\
					&	2	&*&	126.02$\pm$0.39	&	2865.06	&	0.91$\pm$0.01&	3.96$\pm$0.01	\\
					&	10	&*&	124.90$\pm$0.63	&	2835.10	&	0.92$\pm$0.01&	3.96$\pm$0.01	\\
					&	100	&*&	125.95$\pm$0.34	&	1963.95	&	0.80$\pm$0.02&	3.94$\pm$0.01	\\
					&	200	&*&	130.24$\pm$0.32	&	1522.35	&	0.57$\pm$0.04&	3.84$\pm$0.01	\\
					&	500	&*&	134.67$\pm$0.38	&	1278.77	&	0.30$\pm$0.04&	3.71$\pm$0.01	\\
					&	800	&*&	131.22$\pm$0.38	&	1116.76	&	0.22$\pm$0.05&	3.82$\pm$0.01	\\
\caption{Dataset I: Simulations for four fixed values of Ra with various values of Ha. When using the slope method to determine $\delta_v$, a linear region $0 < z < z_{v,linear} = 0.005$ is used for cases without $*$, and $z_{v,linear} = 0.002$ is used for cases with $*$.}
\end{tabular}}
\end{table}

\begin{table}[H]
\centering

\label{simulation table2}
\resizebox{0.9\textwidth}{!}{\begin{tabular}{llll|llll}
\hline
$Ha$		&	$Ra$ 			&   $Nu$ 			& $Re$	&$Ha$		&	$Ra$ &   $Nu$	&	$Re$\\
\hline 
10 & 5 $\times 10^5$ 	& 6.49$\pm$0.05  & 12.40	&  200  & 5 $\times 10^6$ 	& 7.95$\pm$0.06  &11.77\\
   & 2 $\times 10^6$  	& 10.24$\pm$0.11 & 29.62	&    	& 2 $\times 10^7$ 	& 15.85$\pm$0.20 &35.40\\
   & 2 $\times 10^7$  	& 19.89$\pm$0.26 & 104.55	&    	& 5 $\times 10^7$ 	& 23.62$\pm$0.23 &67.25\\
   & 5 $\times 10^7$  	& 25.91$\pm$0.25 & 173.83	&    	& 2 $\times 10^8$ 	& 39.47$\pm$0.43 &174.75\\
   & 2 $\times 10^8$  	& 39.22$\pm$0.39 & 369.74	&    	& 5 $\times 10^8$ 	& 53.50$\pm$0.40 &276.30\\
   & 5 $\times 10^8$  	& 51.14$\pm$0.40 & 597.31	&    	& 2 $\times 10^9$ 	& 82.13$\pm$0.68 &669.50\\
   & 2 $\times 10^9$  	& 77.21$\pm$0.64 & 1248.85	& & & &	\\
                                    \\     
100& 5 $\times 10^5$	& 4.45$\pm$0.01  & 4.14		&  500 	&  2 $\times 10^7$ 	& 8.75$\pm$0.03 &13.64\\
   & 1 $\times 10^6$ 	& 5.94$\pm$0.02  & 6.76		&    	&  5 $\times 10^7$ 	& 14.70$\pm$0.10 &29.88\\
   & 2 $\times 10^6$ 	& 8.13$\pm$0.04	 & 11.54	&    	&  2 $\times 10^8$ 	& 26.25$\pm$0.25 &88.06\\
   & 5 $\times 10^6$	& 11.43$\pm$0.05 & 20.52	&     	&  5 $\times 10^8$  & 45.30$\pm$0.31 &139.24\\
   & 2 $\times 10^7$	& 19.21$\pm$0.23 & 51.65	& & & & \\
   & 5 $\times 10^7$	& 26.52$\pm$0.26 & 98.11	& & & & \\
   & 2 $\times 10^8$	& 40.78$\pm$0.40 & 220.00	& & & & \\
   & 5 $\times 10^8$	& 53.07$\pm$0.29 & 376.99	& & & & \\

\caption{Dataset II: Simulations for four fixed values of Ha with various values of Ra. }
\end{tabular}}
\end{table}

\end{document}